\documentclass[5p,authoryear,times,twocolumn]{elsarticle}
\usepackage{natbib}
\usepackage{color}
\usepackage[latin1]{inputenc}
\usepackage{graphics}
\usepackage{booktabs}
\usepackage{amsmath}
\usepackage{amssymb}
\usepackage{overpic}
\usepackage{booktabs}
\journal{Icarus}
\begin{document}
\begin{frontmatter}
\title{Outgassing of icy bodies in the Solar System - II. Heat transport in dry, porous surface dust layers}
\author[igep]{B. Gundlach}
\ead{b.gundlach@tu-bs.de}
\author[igep]{J. Blum}
\address[igep]{Institut für Geophysik und extraterrestrische Physik, Technische Universität Braunschweig, \\Mendelssohnstr. 3, D-38106 Braunschweig, Germany}
\begin{abstract}
In this work, we present a new model for the heat conductivity of porous dust layers in vacuum, based on an existing solution of the heat transfer equation of single spheres in contact. This model is capable of distinguishing between two different types of dust layers: dust layers composed of single particles (simple model) and dust layers consisting of individual aggregates (complex model). Additionally, we describe laboratory experiments, which were used to measure the heat conductivity of porous dust layers, in order to test the model. We found that the model predictions are in an excellent agreement with the experimental results, if we include radiative heat transport in the model. This implies that radiation plays an important role for the heat transport in porous materials. Furthermore, the influence of this new model on the Hertz factor are demonstrated and the implications of this new model on the modeling of cometary activity are discussed. Finally, the limitations of this new model are critically reviewed.
\end{abstract}
\end{frontmatter}

\setlength{\tabcolsep}{10pt}
\renewcommand{\arraystretch}{2}
\renewcommand{\topfraction}{1.0}
\renewcommand{\bottomfraction}{1.0}
\section{Introduction}
Small bodies in the Solar System are either covered by regolith, a granular medium formed through the continuous bombardment by interplanetary bodies (asteroids) or even consist entirely of granular material (comet nuclei). The heat conductivity of the surface layers of such bodies plays an important role for the minimum and maximum temperature as well as for the temporal temperature evolution on the surfaces of these bodies. It also determines the amount of energy transported into deeper regions of these bodies and, thus, the temperature stratification. In the young Solar System, the fate of planetesimals with respect to sintering and melting is also determined by the heat conductivity of the granular material they consist of \citep[see, e. g., ][]{HeveySanders2006, Gupta2007, MoskovitzGaidos2011, Krause2011}. For comet nuclei, the heat conductivity of the ice-free surface layers determines (among other factors) the temperature and, thus, the evaporation rate of the underlying volatiles.
\par
In paper I \citep{Gundlach2011}, we presented measurements of the gas transport through porous dust layers and derived the gas permeability for arbitrary layers of (spherical and monodisperse) $\mathrm{SiO_2}$ dust particles. In this paper (paper II), we describe laboratory experiments on the determination of the heat conductivity of granular dust layers consisting of spherical $\mathrm{SiO_2}$ dust particles of different sizes (Sect. \ref{Experimental technique} - \ref{Results}). Thereafter (Sect. \ref{Theoretical}), we adopt a model for the heat conductivity of packed spherical granular particles by accounting for the mutual van der Waals interaction of the grains and extend the model to arbitrary granular packing of dust aggregates, as hypothesized to exist on the surfaces of comet nuclei \citep{SkorovBlum2011}. In Sect. \ref{Discussion}, we discuss the implications of this model on the Hertz factor and on the modeling of cometary activity. Furthermore, the limitations of this new model are critically reviewed. Finally, we conclude this paper by critically addressing the capabilities of the heat-conductivity model for future applications in planetary science (Sect. \ref{Conclusion}).

\section{Experimental technique}\label{Experimental technique}

\subsection{Setup}
\begin{figure}[t]
\centering
\includegraphics[angle=0,width=0.5\textwidth]{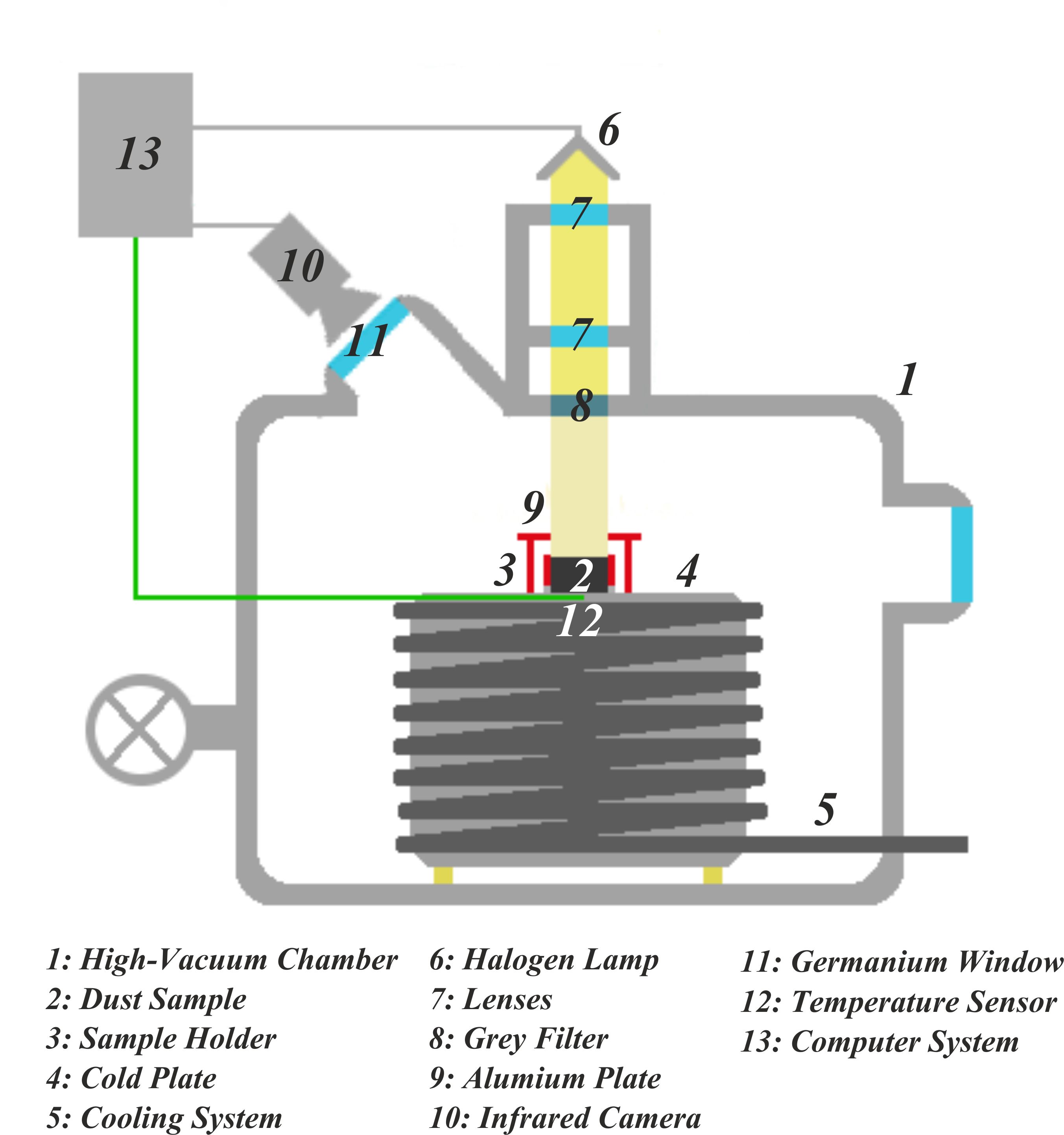}
\caption{Schematic diagram of the adapted experimental setup. See \citet{Gundlach2011} for the unmodified version.}
\label{Aufbau}
\end{figure}

The experiments were performed using the experimental setup introduced in paper I (see Fig. \ref{Aufbau}). Minor modifications of the setup were conducted in order to measure the surface temperature and, therewith, the heat conductivity of porous dust layers.
\par
The measurements were performed in a high-vacuum chamber (see no. 1 in Fig. \ref{Aufbau}) at pressures below $10^{-4}\,\mathrm{mbar}$. In this case, the mean free path of the gas molecules is much larger than the mean pore size of the dust layer. Thus, heat conduction due to gas diffusion can be neglected in the experiments.
\par
The dust samples (2; see Sect. \ref{The dust samples}) were filled into a cylindrical sample holder (3), which was positioned on the cold plate (4). In paper I, the cold plate was cooled by the cooling system (5), which was not used in this work. Here, the cold plate acts as a constant-temperature reservoir.
\par
The dust samples were irradiated with a halogen lamp (6; intensity: $5.32$ solar constants on the surface of the sample). The beam profile of the light source was adjusted with two lenses (7). Different grey filters (8) with transmission rates of $50 \, \%$, $20 \, \%$, $10 \, \%$ and $5 \, \%$, were introduced into the light beam to vary the intensity of the irradiation, without changing the spectrum of the light source.
\par
An aluminum ring (9) was implemented on top of the sample holder, to avoid additional heating of the sample holder and, therewith, additional heating of the dust samples. In order to minimize the heat flow through the sample holder, a material with a low specific heat conductivity (for comparison, metals, glass or plastic materials typically have specific heat conductivities of $\lambda \, = \,  10 \,  - \, 400 \, \mathrm{W\,K^{-1}\,m^{-1}}$, $\lambda \,=  \, 1 \, - \, 3 \mathrm{W\,K^{-1}\,m^{-1}}$ and $\lambda \, = \, 0.2 \, - \,  0.5 \,  \mathrm{W\,K^{-1}\,m^{-1}}$) was chosen (polyvinyl chloride, $\lambda \, = \, 0.15 \, \mathrm{W\,K^{-1}\,m^{-1}}$) and the thickness of the wall of the sample holder was small ($2\,\mathrm{mm}$) in comparison with the diameter of the dust samples ($25\,\mathrm{mm}$) so that the amount of heat transported through the sample holder is much smaller than the heat transported through the dust sample. Furthermore, the sample holder was in a very loose contact with the cold plate, which further decreased the heat flow through the sample holder.
\par
The surface temperature of the dust layer was measured with an infrared camera (10) through a germanium window (11). Due to the low transmissivity of germanium in the infrared (transmission coefficient: $\sim0.4$), a calibration of the infrared camera observing the surface of the dust sample through the germanium window was necessary (see Sect. \ref{Infrared-camera calibration}). A temperature sensor (12) was incorporated into the cold plate, directly beneath the dust sample (distance between the bottom of the dust sample and the temperature sensor: $2\,\mathrm{mm}$). With this sensor, the temperature evolution of the cold plate and, therefore, the temperature evolution of the bottom of the dust sample was monitored. The temperature of the cold plate remained constant during the experiments, due to the high heat capacity of the material (see Fig. \ref{Example} in Sect. \ref{Results}). To control the experiment, a computer system (13) was used.

\subsection{Calibration of the infrared camera}\label{Infrared-camera calibration}
Observing the surface of the dust sample with the infrared camera through a germanium window leads to a falsification of the observed temperature. Thus, a calibration of the infrared camera observing the surface of the sample through the germanium window was necessary. Therefore, the temperature evolution of a stove top was measured with and without germanium window. From this calibration measurement, the influence of the germanium window on the measured temperature was derived. The difference between the measured temperature with and without germanium window was used to calculate the real surface temperature of the dust sample.

\subsection{Procedure}\label{Procedure}
At the beginning of the experiments, the dust samples were filled into the sample holder on top of the cold plate. Then, the experimental chamber was evacuated to low pressures ($p \, < \, 10^{-4}\, \mathrm{mbar}$), so that the mean free path of the gas molecules was much larger than the pore size of the dust sample. In this case, heat transport due to the gas diffusion was negligible. After that, the dust samples were irradiated by the light source until the equilibrium temperature was reached (see Fig. \ref{Example} in see Sect. \ref{Results}). The temperature evolution of the surface of the dust sample and the bottom of the dust sample were measured by the infrared camera and the temperature sensor, respectively.

\section{The dust samples}\label{The dust samples}
\begin{figure*}[!t]
\centering
\includegraphics[angle=0,width=1.0\textwidth]{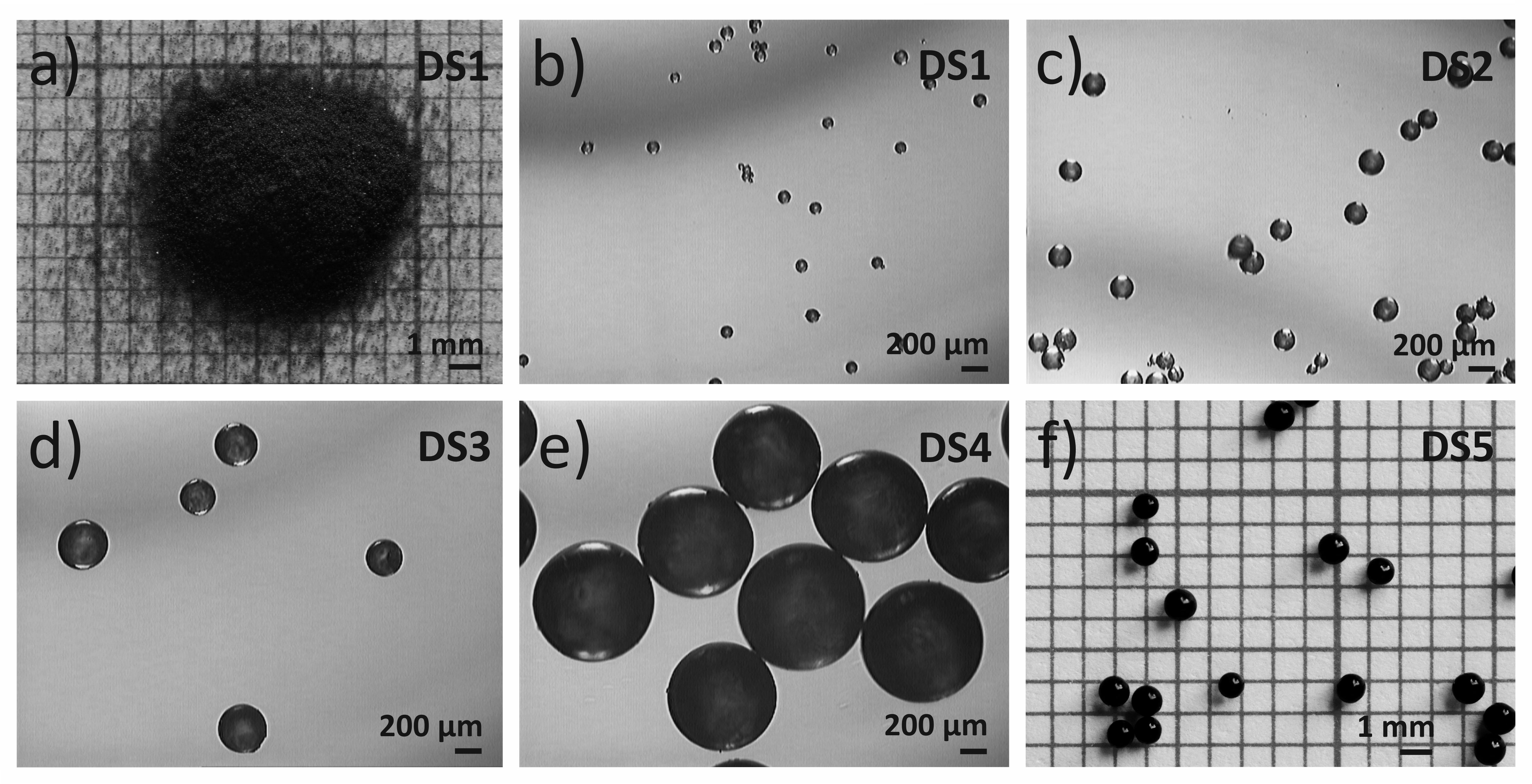}
\caption{Images of a black dust sample (a; DS1) and of the individual $\mathrm{SiO_2}$ particles: DS1 (b), DS2 (c), DS3 (d), DS4 (e) and DS5 (f). The images a and f were taken with a digital camera and the images b - e were taken using a light microscope.}
\label{staub_bild}
\end{figure*}

Five different dust samples (DS1, DS2, DS3, DS4 and DS5) were used in the experiments. All samples consisted of spherical, black $\mathrm{SiO_2}$ particles (see Fig. \ref{staub_bild}). In this section, we discuss the physical properties of the dust samples relevant for this work, such as the size distributions of the particles, the volume filling factors of the dust samples, the plane albedo and the emissivity of the dust samples.

\subsection{Size distribution and volume filling factor}\label{Size distribution and volume filling factor}
\begin{figure}[!t]
\centering
\includegraphics[angle=180,width=0.5\textwidth]{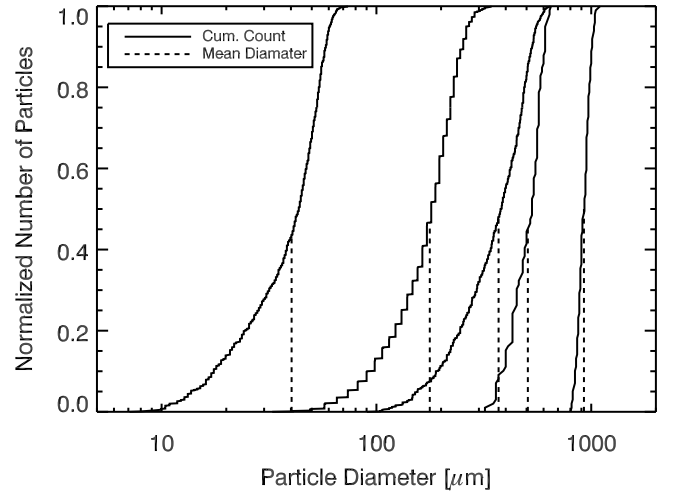}
\caption{Particle size distributions for the five different dust samples (solid curves). The mean particle diameters are visualized by the dashed lines: $s \, = \, (40 \pm 15)\,\mathrm{\mu m}$ (DS1), $s \, = \,(177 \pm 59)\,\mathrm{\mu m}$ (DS2), $s \, = \,(370 \pm 126)\,\mathrm{\mu m}$ (DS3), $s \, = \,(507 \pm 84)\,\mathrm{\mu m}$ (DS4) and $s \, = \,(925 \pm 64))\,\mathrm{\mu m}$ (DS5).}
\label{hist}
\end{figure}

For the estimation of the particle diameters $s$, we used a light microscope (DS1, DS2, DS3 and DS4) and a calliper (DS5). Fig. \ref{hist} shows the resulting particle diameters. The volume filling factors $\phi$ of the dust samples were measured by comparing the mass of the dust sample with the mass of solid $\mathrm{SiO_2}$, which would occupy the same volume.
\par
Tab. \ref{Table_dust} summarizes the mean particle diameters and the volume filling factors of the five different dust samples.
\begin{table}[h!]
\begin{center}
    \footnotesize
    \caption{Comparison of the mean particle diameters $s$ and the volume filling factors $\phi$ of the five dust samples (DS1, DS2, DS3, DS4 and DS5). The errors of the mean particle diameters and the volume filling factors are given by the standard deviations of the measurements.}\vspace{1mm}
    \begin{tabular}{lccc}
        \bottomrule
        Dust Sample & $s$ [$\mathrm{\mu m}$] & $\phi$  \\
        \midrule
        DS1 &  $ \hspace{0.17cm}  40 \pm 15$       & $0.670 \pm 0.015$          \\
        DS2 &  $  177 \pm 59$     & $0.634 \pm 0.024$           \\
        DS3 &  $ \hspace{0.15cm}371 \pm 126$    & $0.688 \pm 0.018$         \\
        DS4 &  $ 507 \pm 84 $    & $0.685 \pm 0.017$            \\
        DS5 &  $ 925 \pm 64 $    & $0.688 \pm 0.008$          \\
        \bottomrule
    \end{tabular}
     \label{Table_dust}
     \end{center}
\end{table}

\subsection{Albedo}
The plane albedo of the dust samples was estimated using the experimental technique invented by \citet{Krause2011}. In this experiment, porous dust samples were positioned inside a vacuum chamber and irradiated by an infrared-laser. The spatial and the temporal temperature evolutions of the surface of the dust sample were monitored with an infrared camera. The heat conductivity and the plane albedo of the dust samples were independently derived by comparing the experimental results with a thermophysical model.
\par
Using this technique, a plane albedo of $A \, = \, 0.35$ was estimated for the dust samples (Krause and Skorov, personal communication).

\subsection{Emissivity}\label{Emissivity}
The determination of the emissivity of the dust sample is important for the calculation of the heat conductivity using the energy balance equation. Thus, the emissivity of the dust samples was measured with the infrared camera. The samples were not heated or cooled so that the temperature of the sample was equal to the ambient temperature, which was monitored by several temperature sensors. For this measurement, the observation angle between the surface of the sample and the infrared camera was $45°$, which is identical to the observation angle in the experiments.
\par
The software of the infrared camera was then used to adjust the emissivity until the infrared camera has measured the correct temperature of the dust sample, assuming Lambertian emission. Using this technique, an emissivity of $\epsilon_{Dust} = 0.99$ was derived. Here, the emissivity and the plane albedo (over all wavelengths) can not be correlated via Kirchhoff's law, because the emissivity was only measured within the spectral range of the infrared camera, limited by the germanium window ($7.5 \, \mathrm{\mu m} \, \leq \, \mathrm{wavelength} \, \leq \, \sim 12 \, \mathrm{\mu m} $).

\section{Experimental results}\label{Results}

\subsection{Surface Temperature}\label{Surface Temperature}
\begin{figure}[t]
\centering
\includegraphics[angle=180,width=0.5\textwidth]{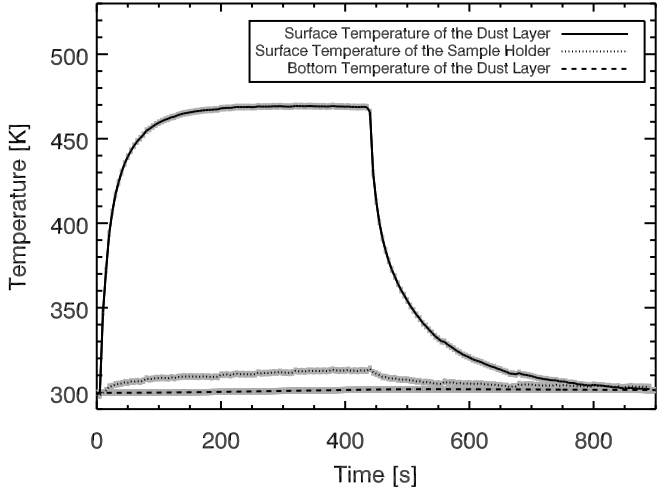}
\caption{Temporal evolution of the surface temperature of the dust layer (solid curve), of the surface temperature of the sample holder (dotted curve) and of the bottom temperature of the dust layer (dashed curve) during and after irradiation ($I \, = \, 5.32 \, I_S$). In this experiment, the DS1 dust sample with a height of $h \, = \,(1.30 \pm 0.01) \, \mathrm{mm}$ was used. The gray regions along the curves represent the measurement uncertainties.}
\label{Example}
\end{figure}
We measured the temporal evolution of the surface temperature of the dust layer, $T_{Surf}$, and of the bottom temperature of the dust layer, $T_{Bottom}$, for different heights of the dust layer, $h$, different particle sizes and different intensities of the light source. The real surface temperature was calculated using the calibration of the infrared camera, which was discussed in Sect. \ref{Infrared-camera calibration} and \ref{Emissivity}.
\par
Fig. \ref{Example} shows the temporal temperature evolutions of the surface of the dust layer (solid curve), of the surface of the sample holder (dotted curve) and of the bottom of the dust layer (dashed curve) during and after irradiation. In this example, the DS1 dust sample with a height of $h \, = \, (1.30 \pm 0.01) \, \mathrm{mm}$ was used. The surface was irradiated with an intensity of $I \, = \, 5.32 \, I_S$ ($I_S$: Solar constant). During all experiments, the bottom temperature of the dust layer was $T_{Bottom} \, = \, \sim 300 \, \mathrm{K}$. Note that the change of the bottom temperature, $\Delta T_{Bottom} \,<\, 2 \, \mathrm{K}$, is small compared to the change of the surface temperature during the experiment. The error on the temperature measurements was $\Delta T_{Surf} \, \pm \, 2.0 \, \mathrm{K}$ (infrared camera) and $\Delta T_{Bottom} \,\pm \, 1.5 \, \mathrm{K}$ (temperature sensor). The errors are visualized by the gray areas in Fig. \ref{Example}.
\par
\begin{figure}[t]
\centering
\includegraphics[angle=180,width=0.5\textwidth]{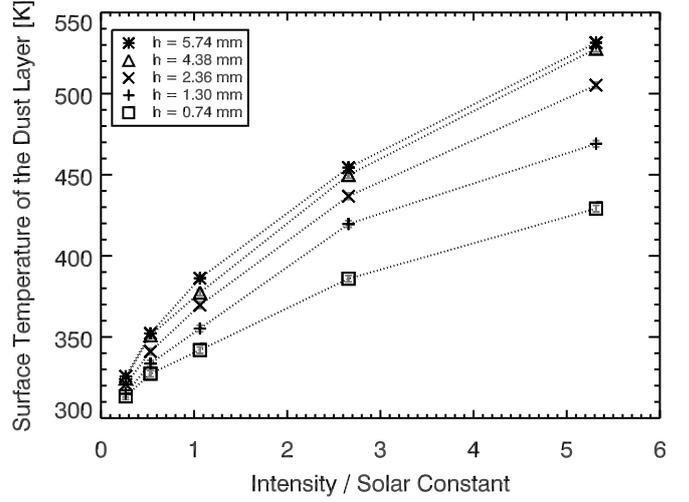}
\caption{Equilibrium surface temperatures of the dust layers as a function of the intensity of the light source. The experiments were carried out for five different dust layer heights, $(0.74 \pm 0.01) \, \mathrm{mm}$ (squares), $(1.30 \pm 0.01) \, \mathrm{mm}$ (pluses), $(2.36 \pm 0.01) \, \mathrm{mm}$ (crosses), $(4.38 \pm 0.01) \, \mathrm{mm}$ (triangles) and $(5.74 \pm 0.01) \, \mathrm{mm}$ (asterisks). The dotted lines are introduced to guide the eye.}
\label{SurfaceTempIntensity}
\end{figure}
\begin{figure}[t]
\centering
\includegraphics[angle=180,width=0.5\textwidth]{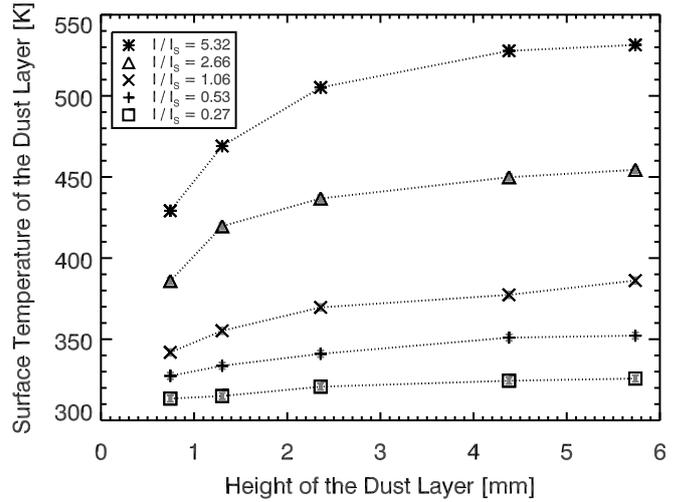}
\caption{Equilibrium surface temperatures of the dust layers with respect to the height of the dust layer for different intensities of the light source: $0.27 \, I_S$ (squares), $0.53 \, I_S$ (pluses), $1.06 \, I_S$ (crosses), $2.66 \, I_S$ (triangles) and $5.32 \, I_S$ (asterisks). The dotted lines are introduced to guide the eye.}
\label{SurfaceTempHeight}
\end{figure}

In total, 25 experiments were carried out using the DS1 dust samples. Five different dust layer heights, $(0.74 \pm 0.01) \, \mathrm{mm}$, $(1.30 \pm 0.01) \, \mathrm{mm}$, $(2.36 \pm 0.01) \, \mathrm{mm}$, $(4.38 \pm 0.01) \, \mathrm{mm}$ and $(5.74 \pm 0.01) \, \mathrm{mm}$, and five different intensities of the light source, $0.27 \, I_S$, $0.53 \, I_S$, $1.06 \, I_S$, $2.66 \, I_S$ and $5.32 \, I_S$, were used. Fig. \ref{SurfaceTempIntensity} and \ref{SurfaceTempHeight} are showing the resulting equilibrium temperatures, obtained in these experiments, with respect to the intensity of the light source and with respect to the height of the dust layer. The dotted lines are implemented to guide the eye.
\par
\begin{figure}[t]
\centering
\includegraphics[angle=180,width=0.5\textwidth]{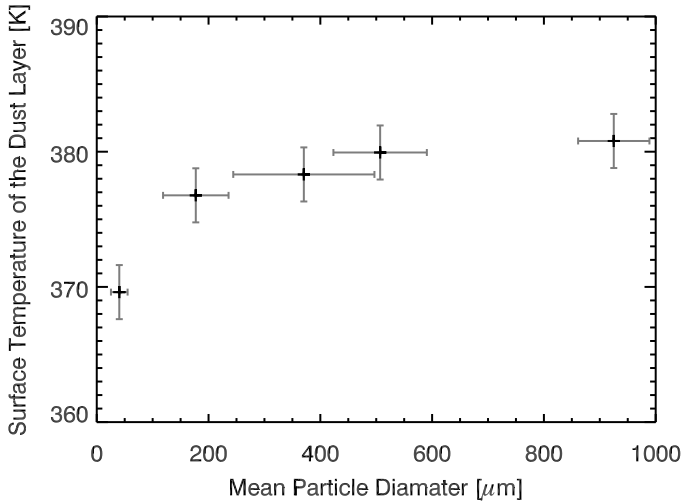}
\caption{Resulting equilibrium surface temperatures of the dust layers for different mean particle diameters. In these experiments, the height of the dust layers and the intensity were kept constant, $h \, = \, (2.36 \pm 0.01) \, \mathrm{mm}$ and $I \, = \, 1.06 \, I_S$, respectively.}
\label{SurfTempDiameter}
\end{figure}

Furthermore, the evolution of the surface temperature of the dust layer was measured for different particles sizes. Fig \ref{SurfTempDiameter} shows that the equilibrium surface temperature of the dust layer increases slightly with the mean diameter of the particles. In these experiments, the height of the dust layers and the intensity were kept constant, $h \, = \, (2.36 \pm 0.01) \, \mathrm{mm}$ and $I \, = \, 1.06 \, I_S$. The mean diameters of the particles were summarized in Sect. \ref{Size distribution and volume filling factor}.

\subsection{Energy balance}
The energy balance equation was used to calculate the heat conduction through the dust layers in thermodynamic equilibrium (constant surface temperature), when the different energy fluxes remained constant. Then, the conducted heat though the dust layer is given by,
\begin{equation}
E_{Cond} \, = \, E_{Irr} \, - \, E_{Rad} \, \mathrm{,}
\label{Fenergybalance1}
\end{equation}
where $E_{Irr} = I \, (1 \, - \, A)$ is the energy input per unit surface into the sample due to irradiation and $E_{Rad} \, = \, \sigma \, ( \, \epsilon_{Dust} \, T^4_{Surf} \, - \, \epsilon_{Back} \, T_{Back}^4)$ is the thermal radiation of the surface of the dust layer, including the thermal radiation of the background. I is the intensity of the light source, $A$ is the plane albedo of the dust sample, $\sigma$ is Boltzmann's constant, $\epsilon_{Dust}$ is the emissivity of the dust sample, $\epsilon_{Back}$ is the emissivity of the background and $T_{Back}$ is the temperature of the background, with $T_{Back} \, = \,  300 \pm 2  \, \mathrm{K}$ (ambient temperature). In this work we assume that the emissivity of the background is unity (black body radiation).
\par
\begin{figure}[t]
\centering
\includegraphics[angle=180,width=0.5\textwidth]{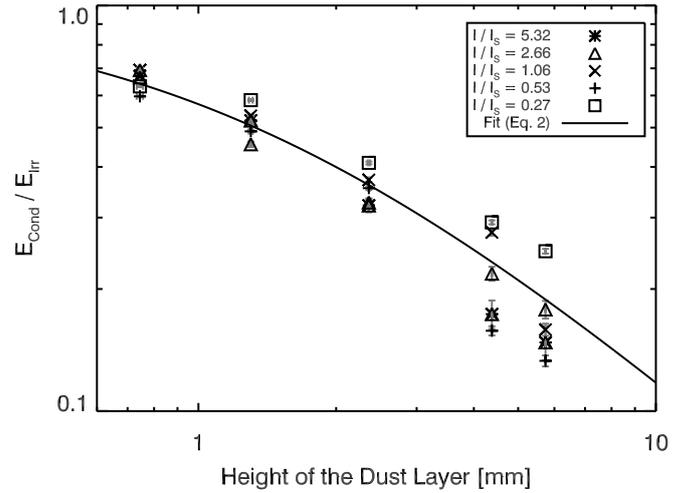}
\caption{Conducted heat through the dust layers, normalized to the energy input into the samples. The heat fluxes were calculated for the 25 experiments, in which the DS1 dust samples were used (asterisks, triangles, crosses, pluses and squares). A good fit to the data (solid curve) is given by Eq. \ref{Fenergybalance2} with $a = (7.41\pm0.29)\,\mathrm{mm^{-1}}$.}
\label{ConductedHeat}
\end{figure}

We calculated the conducted heat through the dust layer for the 25 experiments, in which the DS1 dust samples were used. Fig. \ref{ConductedHeat} shows the conducted heat, normalized to the energy input into the sample, with respect to the height of the dust layer. The normalized conducted heat decreases with the height of the dust layer, which can easily be explained with the decrease of the temperature gradient inside the layer. The temperature gradient decreases linearly with the height of the dust layer, if the surface temperature and the bottom temperature of the dust layer are fixed. If the surface temperature increases with increasing height of the dust layer, the decrease of the temperature gradient is smaller compared with the case discussed above. This behavior can be found in Fig. \ref{ConductedHeat}, because the maximum surface temperatures of the dust layers were still slightly increasing with increasing heights of the dust layers (see Fig. \ref{SurfaceTempHeight}). However, a transition to the linear decrease of the normalized conducted heat through the dust layer can also be seen (see Fig. \ref{ConductedHeat}). Thus, a reasonable fit to the data (solid curve) is given by the following function,
\begin{equation}
\frac{E_{Cond}(h)}{E_{Irr}} \, = \, \frac{1}{ 1 \, + \, a^{-1} \, h } \, \mathrm{,}
\label{Fenergybalance2}
\end{equation}
with $a = (1.35\pm0.05)\,\mathrm{mm}$.
\par
The error due to heat conduction through the sample holder can be neglected, because of the shallow temperature increase of the surface of the sample holder (see Fig. \ref{Example}) and because of the small wall thickness of the sample holder compared with the diameter of the dust layer. Thus, the uncertainty of the derived heat flux trough the dust layer is mostly given by the uncertainty of the temperature measurement.

\subsection{Heat conductivity}\label{Heat conductivity}
\begin{figure}[t]
\centering
\includegraphics[angle=0,width=0.5\textwidth]{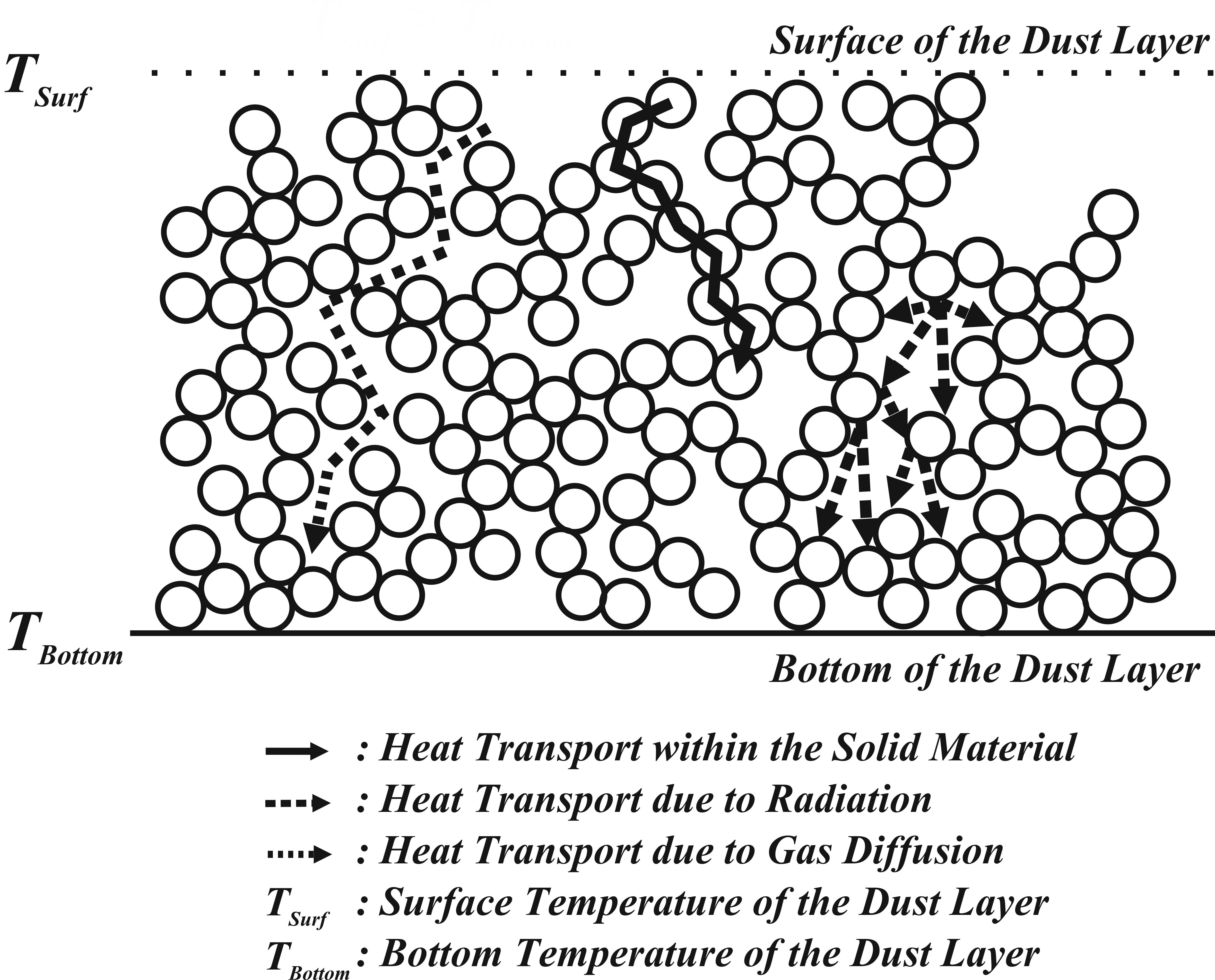}
\caption{Sketch of the heat transport in a porous dust layer. Energy can be transported inside the solid material (solid arrow), due to radiation inside the pores of the material (dashed arrow) and due to gas diffusion (dotted arrow). In this example, all energy fluxes are directed downwards, because the surface temperature of the dust layer is higher than the temperature of the bottom.}
\label{DustLayer}
\end{figure}

In porous materials, heat can be transported inside the solid material, due to radiation inside the pores of the material and due to gas diffusion (see Fig. \ref{DustLayer}).
The total heat conductivity of porous materials is a then given by the combination of these three effects,
\begin{equation}
\lambda(r,T,\Lambda(r)) \, = \, \lambda_{Solid}(r,T) \, + \, \lambda_{Rad}(T,\Lambda(r)) \, + \, \lambda_{Gas} \, \mathrm{.}
\label{HeatCond1}
\end{equation}
Here, $\lambda_{Solid}(r,T)$ is the heat conductivity through the solid network of particles and $r$ is the radius of the particles. The heat conductivity of a porous material can be orders of magnitudes lower than the heat conductivity of the bulk material, due to the small contact areas between the particles, which are determining the efficiency of the heat exchange between the particles. This effect is often taken into account by the introduction of the so-called Hertz factor $H(r,T)$, $\lambda_{Solid}(r,T) \, = \, H(r,T) \, \lambda_{Bulk}(T)$ (see Sect. \ref{Influence on the Hertz factor} for details), where $\lambda_{Bulk}(T)$ is the heat conductivity of the bulk material. For vitreous $\mathrm{SiO_2}$, the temperature dependence of the bulk heat conductivity can be described with a linear function,
\begin{equation}
\lambda_{Bulk,SiO_2}(T) \, = \, b_1 \,T \, + \, b_2 \, \mathrm{,}
\label{HeatCond1.1}
\end{equation}
with $b_1 \, = \, (1.26\,\pm\,0.05)\times10^{-3}\,\mathrm{W\, K^{-2} \, m^{-1} }$ and $b_2 \, = \, (9.94\,\pm\,0.20)\times10^{-1}\,\mathrm{W \, K^{-1} \, m^{-1}}$  \citep{Ratcliffe1963}.
\par
The radiative heat transport in a porous material can be calculated treating the radiation as a photon gas \citep{Merrill1969}, or by the formulation introduced by \citet{Schotte1960}. Thus, the heat conductivity of the radiative transport can be written as,
\begin{align}
\lambda_{Rad}(T,\Lambda(r)) \, &= \, \kappa \, T^3 \, \Lambda(r)
\label{HeatCond2}
\end{align}
with
\begin{align}
\kappa \, = \, \begin{cases} 16 \, / \, 3 \,  \sigma \ \ \ \ \ \ \ \ \mathrm{(Merrill,\ 1969)} \\   \ \ \ \, \, 8 \ \hspace{2mm}  \sigma \, \epsilon \  \ \ \ \ \, \mathrm{(Schotte,\ 1960)}  \end{cases} \mathrm{,} \nonumber
\end{align}
where $\Lambda(r)$ is the mean free path of the photons. Note that the first of the two formulations of the radiative heat conductivity does not depend on the emissivity of the dust layer. In this work, we used Eq. \ref{HeatCond2} with $\kappa \, = \, 16 \, / \, 3 \, \sigma$. In our case ($\epsilon_{Dust} \, = \, 0.99$, see Sect. \ref{Emissivity}), both formulations of the radiative heat conductivity differ only by a factor $2 \, / \, 3$, which we regard as negligible. However, for the calculation of the radiative heat conductivity for materials with an emissivity strongly deviating from unity, the second formulation of the radiative heat conductivity should be used.
\par
Furthermore, we can neglect heat transport due to gas diffusion $\lambda_{Gas} \, = \, 0$, because of the low pressures during the experiments (see Sect. \ref{Procedure}).
\par
We calculated the heat conductivity of the dust samples assuming a constant vertical temperature gradient inside the dust layers in the thermodynamic equilibrium. This assumption only holds if the heat conductivity is not dominated by the radiative heat transport (see Sect. \ref{Influence on the modeling of cometary activity}). Here, we used relatively compact samples ($\phi \, = \, 0.63 \, - \, 0.69$, see Sect. \ref{Size distribution and volume filling factor}), where the radiative heat transport is present (see Sect. \ref{A general model for the heat conductivity of dust layers}), but not dominating the heat transport. In this case, the heat conductivity can be calculated as follows,
\begin{equation}
\lambda(r,T,\Lambda(r)) \, = \, E_{Cond}(h) \, \frac{h}{T_{Surf} \, - \, T_{Bottom}} \, \mathrm{.}
\label{HeatCond3}
\end{equation}
\par
\begin{figure}[t]
\centering
\includegraphics[angle=180,width=0.5\textwidth]{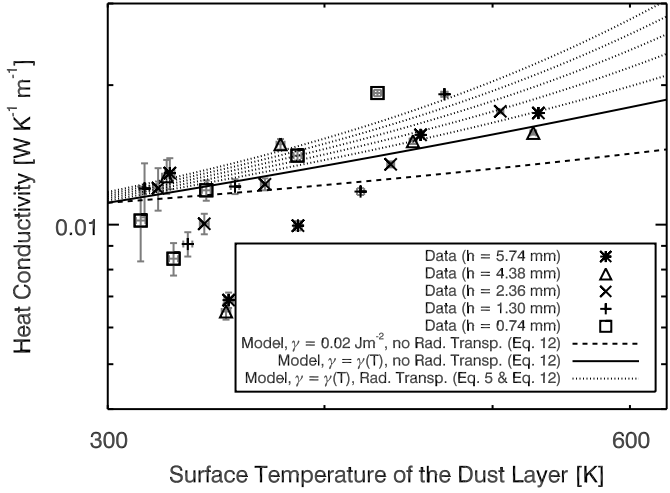}
\caption{Temperature dependence of the heat conductivity (DS1 dust samples; diameter: $(40\pm 15) \, \mathrm{\mu m}$), measured for different dust layer heights: $(0.74 \pm 0.01) \, \mathrm{mm}$ (squares), $(1.30 \pm 0.01) \, \mathrm{mm}$ (pluses), $(2.36 \pm 0.01) \, \mathrm{mm}$ (crosses), $(4.38 \pm 0.01) \, \mathrm{mm}$ (triangles) and $(5.74 \pm 0.01) \, \mathrm{mm}$ (asterisks). The dashed curve shows the temperature dependence of the heat conductivity calculated with the modified model of \citet[][see Sect. \ref{A general model for the heat conductivity of dust layers}, Eq. \ref{Theory3}]{ChanTien1973} assuming that the specific surface energy of $\mathrm{SiO_2}$ is not temperature dependent, $\gamma_{SiO_2} \, = \, 0.02 \, \mathrm{J \, m^{-2}}$. The model predictions for a temperature dependent specific surface energy (see Eq. \ref{Theory2.1}) are visualized by the solid curve. Additionally, the influence of the radiative heat transport (see Eq. \ref{HeatCond2}) is shown for different values of the mean free path, $0.5\, r$, $1.0\,r$, $1.5\,r$, $2.0\,r$ and $2.5\,r$, by the dotted curves. For the theoretical derivations of the heat conductivities, the temperature inside the dust layer was averaged, $T \, = \, T_{Bottom} \, + \, ( \, T_{Surf} \,  -\, T_{Bottom}) \,/ \, 2$ (see Sect. \ref{Heat conductivity} for explanation)}
\label{HeatCondSurfTemp}
\end{figure}

Fig. \ref{HeatCondSurfTemp} shows the derived heat conductivities of the DS1 dust samples (see also Fig. \ref{SurfaceTempIntensity} and Fig. \ref{SurfaceTempHeight}) with respect to the surface temperature of the dust layers. The heat conductivity increases with the surface temperature of the dust layer. The uncertainty of the calculated heat conductivities is primarily given by the error of the temperature measurement. Due to the usage of Eq. \ref{Fenergybalance1}, the derived heat conductivity is calculated from the difference of two large numbers. As a result, the scatter of the derived heat conductivities is bigger for lower temperatures ($T_{Surf} \, \gtrsim \, T_{Back}$) than for higher temperatures ($T_{Surf} \, > \, T_{Back}$). For this reason, we have not included the experiments performed with grey filters having very low transmission rates of $1 \, \%$ and $2 \, \%$. In Sect. \ref{A general model for the heat conductivity of dust layers}, the derived heat conductivities are compared with a general model for the heat conductivity of porous dust layers.
\par
\begin{figure}[t]
\centering
\includegraphics[angle=180,width=0.5\textwidth]{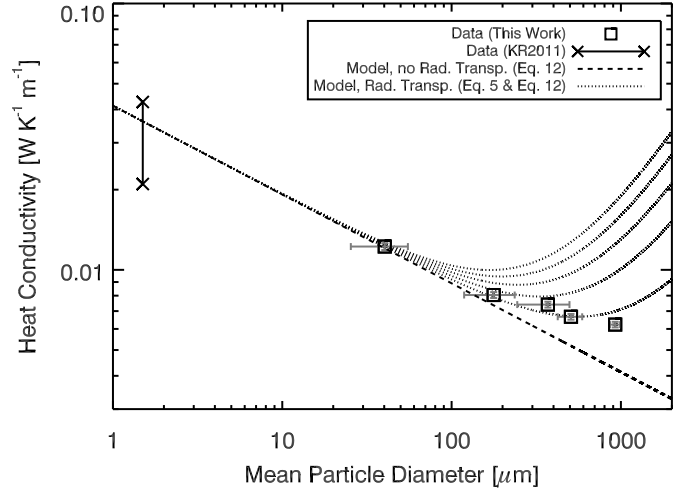}
\caption{Heat conductivity as a function of the mean particle diameter. The experimental results are shown by the squares. In all experiments, the height of the dust layer was identical, $h \, = \, (2.36 \pm 0.01) \, \mathrm{mm}$. For comparison, the results of the model for the heat conductivity of packed spheres with $\phi \, = \, 0.68$ in vacuum (see Eq. \ref{Theory3}) are visualized without the influence of radiative heat transport (dashed curve) and with the influence of radiative heat transport (dotted curves; see Eq. \ref{HeatCond2}). The calculations were performed for a mean temperature of the dust layer of $340 \, \mathrm{K}$ and for five different values of the mean free path of the photons inside the pores of the porous material, $0.5\, r$, $1.0\,r$, $1.5\,r$, $2.0\,r$ and $2.5\,r$. The crosses are experimental results from \citet{Krause2011} for dust layers with packing densities of $\phi \, = \, 0.54$ and $\phi \, = \, 0.67$.}
\label{HeatCondPartSize}
\end{figure}

Furthermore, the influence of the grain size on the surface temperature of the dust layers and, therewith, on the heat conductivity was investigated (see Fig. \ref{SurfTempDiameter}). Therefore, the height of the dust layers was kept constant, $h \, = \, (2.36 \pm 0.01) \, \mathrm{mm}$. Fig. \ref{HeatCondPartSize} shows the derived heat conductivities (squares) with respect to the mean particle diameters. The heat conductivity decreases with increasing mean particle diameter. Note that the heat conductivity is a combination of the heat conductivity of the solid material and the heat conductivity of the radiative transport.
\par
Additionally, the heat conductivities of porous dust samples composed of spherical, $1.5$ micrometer-sized $\mathrm{SiO_2}$ particles \citep{Krause2011} were implemented in Fig. \ref{HeatCondPartSize} (crosses). The lower value, $\lambda \, = \, 0.021 \, \mathrm{W \, K^{-1} \, m^{-1}} $, denotes the heat conductivity of the dust layer with the highest volume filling factor, $\phi \, = \, 0.54 $, measured by \citet{Krause2011}. The upper value, $\lambda \, = \, 0.043 \, \mathrm{W \, K^{-1} \, m^{-1}} $, was derived for $\phi \, = \, 0.67 $, using the correlation between the heat conductivity and the volume filling factor of the material \citep{Krause2011},
\begin{equation}
\lambda(\phi) \, = \, c_1 \, \mathrm{exp}\left[\, c_2 \, \phi\,\right]  \, \mathrm{,}
\label{HeatCond4}
\end{equation}
with $c_1 \, = \, (1.31 \pm 0.48) \times 10^{-3} \, \mathrm{W \, K^{-1} \, m^{-1}}$ and $ c_2 \, = \, 5.19 \pm 0.73$ (Krause, personal communication). These fit parameters were obtained by fitting the data obtained by \citet{Krause2011} without using the value for fused silica.
\par
A detailed theoretical discussion of the heat conductivity of porous dust layers is given in Sect. \ref{Theoretical}, where a general model for the conductivity of dust layers will be presented and compared with the experimental results.

\section{A model for the heat conductivity in loose granular media}\label{Theoretical}
\subsection{A general model for the heat conductivity of dust layers}\label{A general model for the heat conductivity of dust layers}
A very detailed analytical study of the heat conductivity of packed spheres in vacuum was carried out by \citet{ChanTien1973}. They solved the heat transfer equation of single spheres and used this result to calculate the conductivity of packed spheres. In their model, heat can be transported from one sphere to an other only through the contact interface, which can be determined using the Hertzian theory of contact \citep{Hertz1881}. They found that the heat conductivity of the packed spheres in vacuum can be calculated as follows,
\begin{align}
\lambda_{Solid}(r,T,\phi) \, &= \,  \lambda_{Bulk}(T) \, \left[\, \frac{3}{4} \,\frac{1 \, - \, \mu^2}{E(T)} \, F(r) \, r \,\right]^{1/3} \xi(r,\phi) \label{Theory1}
\end{align}
with
\begin{align}
\xi(r,\phi) \, &= \, \frac{1}{0.531 \, S(\phi)} \, \frac{N_A(r)}{N_{L}(r)} \, \mathrm{.}
\label{Theory1a}
\end{align}
Here, $\mu$ and $E(T)$ are the Poisson's ratio and Young's modulus of the material, respectively. $S(\phi)$ is a model parameter, which depends on the packing structure of the material. $N_A(r)$ and $N_L(r)$ are the number of particles per unit area and unit length, respectively. Tab. \ref{Table_packing} summarizes the constants $S(\phi)$, $N_A(r)$ and $N_L(r)$ for three different compact packing structures, simple cubic (sc), body-centered cubic (bcc) and face-centered cubic (fcc).
\begin{table}[b]
\begin{center}
    \scriptsize   
    \caption{Summarization of the constants $S$, $N_A(r)$, $N_L(r)$ and the volume filling factors $\phi$ of the three different packing structures, simple cubic (sc), body-centered cubic (bcc) and face-centered cubic (fcc). See \citet{ChanTien1973} for details.}\vspace{1mm}
    \begin{tabular}{lccc}
        \bottomrule
        Parameter & SC & BCC & FCC  \\
        \midrule
        $S(\phi)$      &  $1$ & $\frac{1}{4}$ & $\frac{1}{3}$  \\
        $N_A(r)$ &  $\frac{1}{4 \, r^2}$ & $\frac{3}{16 \, r^2}$ & $\frac{1}{2\, (3)^{1/2} r^2}$ \\
        $N_L(r)$ &  $\frac{1}{2 \, r}$ & $\frac{3^{1/2}}{2\,r}$ & $\frac{(3/8)^{1/2}}{r}$  \\
        $\phi$   &  $0.52$ & $0.68$ & $0.74$  \\
        \bottomrule
    \end{tabular}
     \label{Table_packing}
     \end{center}
\end{table}
\par
$F(r)$ describes the external force, which acts on the spheres and determines the contact interface between the particles. \citet{ChanTien1973} used an external applied load or the weight of the spheres above the contact to calculate the contact interface. However, the adhesive bonding between two spherical particles in contact (van der Waals bonding) can be orders of magnitude higher than their weight, when no external load is applied. The adhesive bonding between two spherical particles in contact can be calculated, using the theory of \citet{JKR1971},
\begin{equation}
F_{vdW}(r,T) \, = \, 3 \, \pi \, \gamma(T) \, r \, \mathrm{,}
\label{Theory2}
\end{equation}
where, $\gamma(T)$ is the specific surface energy of the material, which is a measure for the adhesive bonding strength of the material. For $\mathrm{SiO_2}$, measurements of the specific surface energy at room temperature yield a mean value of $\gamma_{SiO_2} \, = \, 0.02 \, \mathrm{J \, m^{-2}}$ \citep{Kendall1987, Heim1999, Gundlach2011b}.
However, the specific surface energy should also depend on temperature, because an increase of the temperature increases the rate of induced dipoles per molecule. Since the influence of the temperature on the specific surface energy is not known in detail, we assume that the specific surface energy increases linearly with temperature,
\begin{equation}
\gamma_{SiO_2}(T) \, = d_1 \, T  \, \mathrm{,}
\label{Theory2.1}
\end{equation}
with $d_1 \, = \, 6.67 \times 10^{-5} \, \mathrm{J \, m^{-2}\, K^{-1}}$. The slope is given by the assumptions that the specific surface energy should be reduced to zero if the temperature tends to zero and that the specific surface energy is $\gamma_{SiO_2} \, = \, 0.02 \, \mathrm{J \, m^{-2}}$ for $T \, = \, 300 \, \mathrm{K}$.
\par
In this work, we use the theory of \citet[][see Eq. \ref{Theory1}]{ChanTien1973} to calculate the heat conductivity of our dust samples, but instead of taking the weight of the particles into account, we use the adhesive bonding force (see Eq. \ref{Theory2}), to determine the contact area between the particles. Thus, the heat conductivity of porous media in vacuum without external load and radiative transport is given by,
\begin{align}
\lambda_{Solid}(r,T,\phi) \, = \, & \lambda_{Bulk}(T) \, \left[\, \frac{9}{4} \,\frac{1 \, - \, \mu^2}{E(T)}  \, \pi \, \gamma(T) \, r^2 \,\right]^{1/3}  \xi(r,\phi) \, \mathrm{.}
\label{Theory3}
\end{align}
The dust samples used in this work are random close packed, $\phi \, = \, 0.63 \, - \, 0.69$ (see Sect. \ref{Size distribution and volume filling factor}). Since the coordination number of a random close packed material is very close to the coordination number of a body-centered cubic packed material \citep{Lagemaat2001,Antwerpen2010}, we use the body-centered cubic packing structure to calculate the heat conductivity of the dust samples (see Tab. \ref{Table_packing}).
\par
The bulk heat conductivity and the specific surface energy of $\mathrm{SiO_2}$ are given by Eq. \ref{HeatCond1.1} and Eq. \ref{Theory2.1}, respectively. For completeness it should be mentioned that Young's modulus of $\mathrm{SiO_2}$ is also temperature dependent \citep{MarxSivertsen1953, SpinnerCleek1960}. However, the variation of Young's modulus due to temperature changes is less than $10\,\%$ for temperatures between $300 \, \mathrm{K}$ and $600 \, \mathrm{K}$, which was the highest surface temperature achieved during the experiments. Therefore, we use a constant value for the Young's modulus and Poisson's ratio of $\mathrm{SiO_2}$, $E_{SiO_2} \, = \, 5.51 \times 10^{10} \, \mathrm{Pa}$ and $\mu_{SiO_2} \, = \, 0.17$ \citep{ChanTien1973}.
\par
Using Eq. \ref{Theory3}, the heat conductivity of packed $\mathrm{SiO_2}$ spheres for different temperatures was calculated and compared with the experimental results (see Fig. \ref{HeatCondSurfTemp}). The heat conductivity was derived assuming a constant (i. e. temperature independent) specific surface energy, $\gamma_{SiO_2} \, = \, 0.02 \, \mathrm{J \, m^{-2}}$ (dashed curve). The model is in a good agreement with the experimental results for temperatures below $\sim 400 \, \mathrm{K}$. But, for higher temperatures, the model predictions deviate from the data.
\par
Including a temperature dependent specific surface energy (see Eq. \ref{Theory2.1}) yields a better agreement between the model (solid curve) and the experimental results ($\chi^2$ decreases from $9.19 \times 10^{-6}$ to $6.43 \times 10^{-6}$). Additionally, the influence of the radiative heat transfer was included (see Eq. \ref{HeatCond2}), using different values for the mean free path of the photons inside the voids of the porous material, $\Lambda(r) = 0.5\, r$ ($\chi^2 \, = \, 6.39 \times 10^{-6}$), $1.0\,r$ ($\chi^2 \, = \, 6.90 \times 10^{-6}$), $1.5\,r$ ($\chi^2 \, = \, 7.94 \times 10^{-6}$), $2.0\,r$ ($\chi^2 \, = \, 9.51 \times 10^{-6}$) and $2.5\,r$ ($\chi^2 \, = \, 1.16 \times 10^{-5}$; dotted curves). For $\Lambda(r) = 0.5\, r$ ($\chi^2 \, = \, 6.39 \times 10^{-6}$), the model predictions are in best agreement with the experimental results.
\par
\begin{figure}[t]
\centering
\includegraphics[angle=180,width=0.5\textwidth]{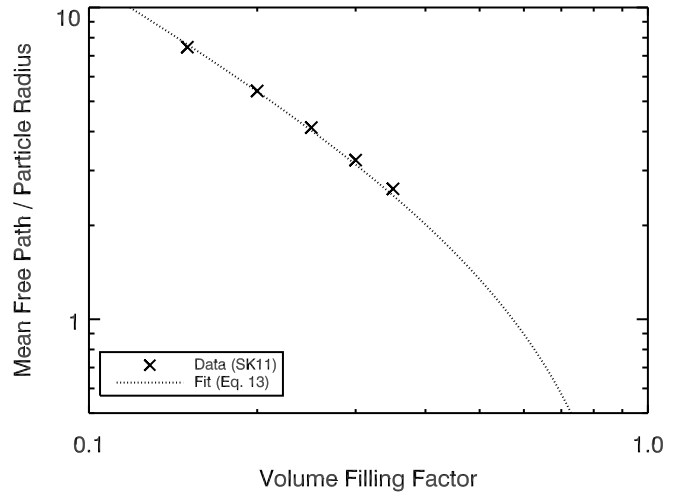}
\caption{Theoretical derivation of the mean free path of the photons inside the voids of porous dust layers for different volume filling factors of the material. The data \citep[crosses; ][]{Skorov2011} were fitted with Eq. \ref{Theory4} (dotted curve) to extrapolate the mean free path to higher volume filling factors.}
\label{MFP}
\end{figure}
The mean free path inside the voids of a porous material was recently investigated theoretically by \citet[][crosses in Fig. \ref{MFP}]{Skorov2011}. They found that the mean free path strongly depends on the volume filling factor of the material. Unfortunately, the highest volume filling factor used in this work, $\phi = 0.35$, is two times lower than the volume filling factor of our dust layers (see Sect. \ref{Size distribution and volume filling factor}). Thus, we used the following fit function, which was theoretical predicted by \citet{Dullien1991}, to extrapolate the mean free path derived by \citet{Skorov2011} to higher volume filling factors (see dotted curve in Fig. \ref{MFP}),
\begin{equation}
\Lambda(r, \phi) \, =  \, e_1 \, \frac{1 \, - \, \phi}{\phi} \,  r \, \mathrm{.}
 \label{Theory4}
\end{equation}
The best fit to the data is given by $e_1 \, = \, 1.34\pm0.01$. Using this formula, we can estimate the mean free path of the photons inside a material with arbitrary volume filling factor. For $\phi = 0.67$, Eq. \ref{Theory4} yields $\Lambda(r) = 0.7 \, r$, which is in a good agreement with the mean free path derived from the comparison of our experimental results with the model for the heat conductivity. Note that our experiments were performed with polydisperse $\mathrm{SiO_2}$ particles (see Fig. \ref{staub_bild} and \ref{hist}) while the calculations were conducted for monodisperse particles. The presence of different particle sizes should influence the mean free path of the photons inside the pores of the material. Another uncertainty is given by the assumption that the specific surface energy increases linearly with the mean temperature of the dust layer (see Eq. \ref{Theory2.1}).
\par
Furthermore, Eq. \ref{Theory3} was used to calculate the influence of the grain size on the heat conductivity of the dust layer. Fig. \ref{HeatCondPartSize} shows a comparison of the derived heat conductivities (dashed curve) with the experimental results (crosses and squares). Note that the heat conductivity of dust samples composed of spherical, monodisperse $1.5$ micrometer-sized particles (crosses) was measured by \citet{Krause2011}. See Sect. \ref{Heat conductivity} for a more detailed description.
\par
The calculations were conducted using a mean temperature of $340 \, \mathrm{K}$ for the dust layer. This value was used because the maximum surface temperature of the dust layers was $\sim 380 \, \mathrm{K}$ (see Fig. \ref{SurfTempDiameter}) and the bottom temperature was $\sim 300 \, \mathrm{K}$ during all experiments (see Fig. \ref{Example}). However, this value can slightly deviate from the real mean temperature of the dust layer, because of the temperature dependence of the heat conductivity, which leads to a non linear temperature gradient inside the dust layer (see Sect. \ref{Influence on the modeling of cometary activity}).
\par
The influence of the radiative heat transport (see Eq. \ref{HeatCond2}) on the model for the heat conductivity is also shown in Fig. \ref{HeatCondPartSize} (dotted curves) for different values for the mean free path of the photons inside the voids of the porous material, $\Lambda(r) = 0.5\, r$, $1.0\,r$, $1.5\,r$, $2.0\,r$ and $2.5\,r$ (dotted curves). A good agreement between the experimental results and the model can be found for $\Lambda(r) = 0.5\, r$, which is consistent with the theoretical prediction of Eq. \ref{Theory4}, $\Lambda(r) = 0.7\, r$. Note that the heat conductivity of the particle network decreases with increasing particle diameter. However, for large particle diameters, the radiative heat transport becomes important, due to the increase of the mean free path of the photons (see Eq. \ref{Theory4}). As a result, the heat conductivity increases with increasing particle diameter. The combination of both behaviors leads to a minimum of the heat conductivity for a certain particle diameter, which depends on the temperature on the material.

\subsection{Heat conductivity of dust layers composed of individual dust aggregates}\label{Heat conductivity of dust layers composed of individual dust aggregates}
\begin{figure}[t]
\centering
\includegraphics[angle=0,width=0.5\textwidth]{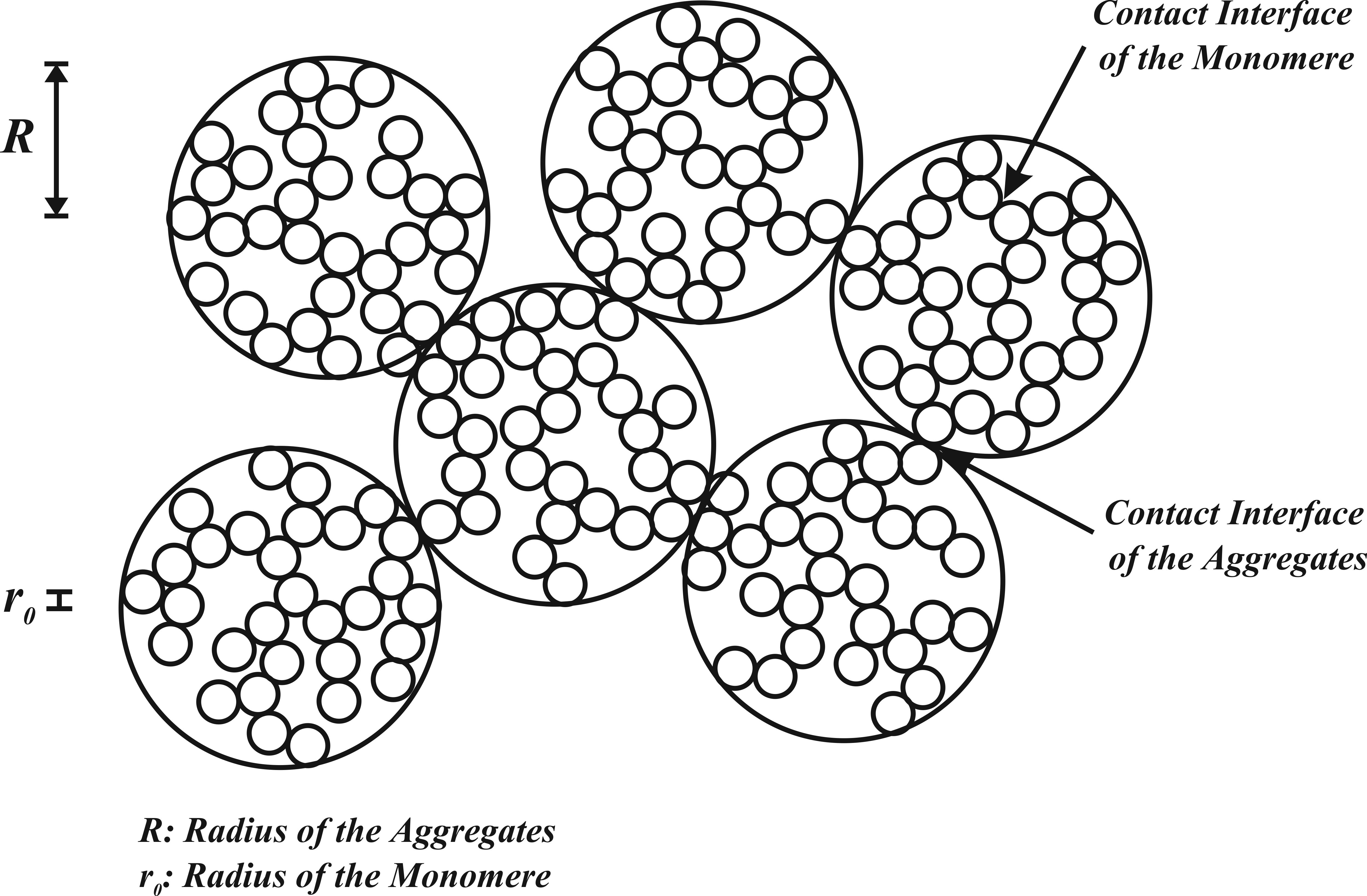}
\caption{Sketch of a dust layer composed of aggregates with radii $R$. The aggregates themselves consist of micrometer-sized dust particles with radii $r_0$. }
\label{Aggregate}
\end{figure}

Dust layers covering the surfaces of planets such as Mars or covering the surfaces of small bodies like comets or asteroids may not only be composed of single grains. It was recently discussed by \citet{SkorovBlum2011} that the non volatile dust layer on the surface of a cometary nucleus could consist of individual dust aggregates, which themselves are composed of micrometer-sized dust particles. These aggregates have formed before they were incorporated into the cometary nucleus. Fig. \ref{Aggregate} shows a sketch of a dust layer composed of dust aggregates. Here, the volume filling factor of the dust layer $\phi_{Layer}$ is a combination of the volume filling factor of the individual dust aggregates $\phi_{Agg}$ and the volume filling factor of the packing structure of the aggregates $\phi_{Struc}$ i. e. $\phi_{Layer} \, = \, \phi_{Agg} \, \phi_{Struc}$.
\par
In this Section, we show how to modify the model derived in Sect. \ref{A general model for the heat conductivity of dust layers} in order to calculate the heat conductivity of dust layers composed of individual dust aggregates. The total heat conductivity of dust layers composed of monodisperse aggregates can be derived by modifying Eq. \ref{Theory3} and by taking the heat conductivity due to radiation into account (see Eq. \ref{HeatCond2}),
\begin{align}
& \hspace{20mm} \lambda_{Layer}(r_0,R,T,\phi_{Struc},\phi_{Agg},\Lambda(R)) \, = \nonumber\\[2mm]
&\lambda_{Agg}(r_0,T,\phi_{Agg}) \times\left[\,\frac{9\, \left( \, 1 \, - \, \mu_{Agg}^2 \, \right)}{4 \, E_{Agg}}  \, \pi \, \gamma_{Agg}(T) \, R^2 \,\right]^{1/3} \, \xi(R,\phi_{Struc}) \nonumber\\[2mm]
& \hspace{26mm}+ \, \lambda_{Rad}(T,\Lambda(R,\phi_{Struc}))  \, \mathrm{.}
\label{Theory5}
\end{align}
$R$, $E_{Agg}$ and $\mu_{Agg}$ are the radius, Young's modulus and Poisson's ratio of the aggregates, respectively. The specific surface energy of the aggregates $\gamma_{Agg}(T)$ can be calculated, using the theory developed by \citet[][their Eqs. 5 - 7]{Weidling2011},
\begin{equation}
\gamma_{Agg}(T) \, = \, \phi_{Agg} \, \gamma_{SiO_2}^{5/3}(T) \, \left[ \frac{9 \, \pi \, \left( \, 1 \, - \mu_{Agg}^2 \, \right)}{r_0 \, E_{SiO_2}} \right]^{2/3}  \, \mathrm{,}
\label{Theory5.1}
\end{equation}
where $r_0$ is the radius of the monomeres.
\par
$\lambda_{Agg}(r_0,T,\phi_{Agg})$ is the heat conductivity of the individual aggregates, which can be also calculated using Eq. \ref{Theory3},
\begin{align}
\lambda_{Agg}(r_0,T,\phi_{Agg}) \, = \, & \lambda_{Bulk}(T) \, \left[\, \frac{9  \,    \left( \, 1 \, - \, \mu_{SiO_2}^2\, \right)}{4 \, E_{SiO_2}}  \, \pi \, \gamma_{SiO_2}(T) \, r_0^2 \,\right]^{1/3} \nonumber\\[2mm] &\times \, \xi(r_0,\phi_{Agg}) \, \mathrm{.}
\label{Theory6}
\end{align}
Note that the volume filling factor of the packing structure of the aggregates $\phi_{Struc}$ can be different to the volume filling factor of the individual aggregates $\phi_{Agg}$.
\par
For simplicity, we assume that the radiative heat transport (second term on the right hand side in Eq. \ref{Theory5}) only takes place in the voids between the aggregates. Therefore, the heat transport inside the aggregates due to radiation is neglected. In this case, the mean free path of the photons can be calculated using Eq. \ref{Theory4} and $\phi = \phi_{Struc}$.
\par
\begin{figure}[t]
\centering
\includegraphics[angle=180,width=0.5\textwidth]{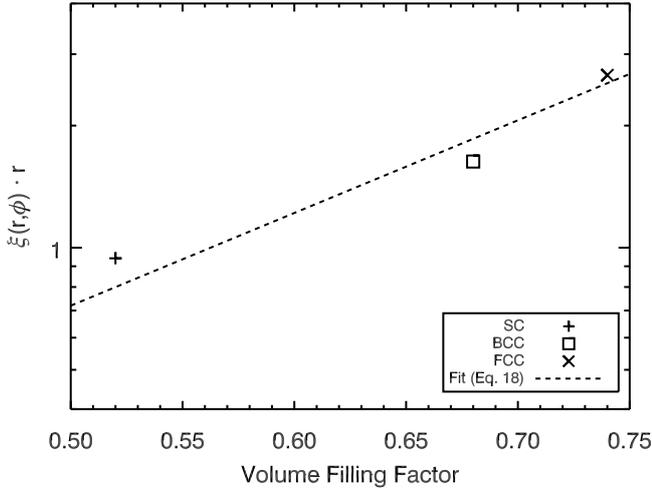}
\caption{Influence of the volume filling factor on the coefficient $\xi(r,\phi) \, r$ (see Eq. \ref{Theory1} and \ref{Theory1a}). The values for the simple cubic structure (sc; plus), the body-centered cubic structure (bcc; square) and the the face-centered cubic structure (fcc; cross) are given by \citet{ChanTien1973}. A good fit to the data is given by Eq. \ref{Theory8} with $f_1 \, = \, (5.18 \pm 3.45) \times 10^{-2}$ and $f_2 \, = \, 5.26 \pm 0.94$.}
\label{Faktor}
\end{figure}

Since dust layers and dust aggregates are not necessarily packed in one of the three close packing structures (see Tab. \ref{Table_packing}), the coefficient $\xi(r,\phi)$ (see Eq. \ref{Theory1a}) has to be modified in order to take other packing structures and, therefore, other volume filling factors of the material into account. The correlation between the coefficient $\xi(r,\phi)$ and the volume filling factor of the material was derived by fitting the dimensionless values (see Tab. \ref{Table_packing} and symbols in Fig. \ref{Faktor})
\begin{equation}
\xi(r,\phi) \, r \, = \, \frac{1}{0.531 \, S(\phi)} \, \frac{N_A(r)}{N_{L}(r)} \, r
\, \mathrm{,}
 \label{Theory7}
\end{equation}
with
\begin{equation}
\xi(r,\phi) \, r \, = \, f_1 \, \mathrm{exp}\Big[ \, f_2 \, \phi \, \Big]
\label{Theory8}
\end{equation}
(dashed curve in Fig. \ref{Faktor}). The fit parameters are $f_1 \, = \, (5.18 \pm 3.45) \times 10^{-2}$ and $f_2 \, = \, 5.26 \pm 0.94$. Extending the model with Eq. \ref{Theory8}, provides the possibility to calculate the heat conductivity of packed spheres for different volume filling factors (i. e. packing structures) of the material.
\par
\begin{figure}[t]
\centering
\includegraphics[angle=180,width=0.5\textwidth]{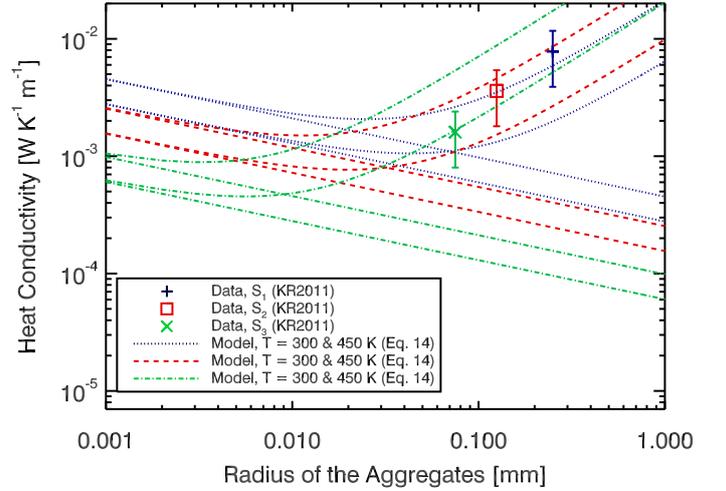}
\caption{Heat conductivity of dust layers composed of individual dust aggregates. The experimental results \citep[][symbols]{Krause2011} are compared with the model for the heat conductivity of dust layers composed of individual aggregates (see Eq. \ref{Theory5}, curves). The calculations were conducted using the parameters corresponding to the three different experiments (see Tab. \ref{Table_Agglomerates}), $S_1$ (blue dotted curves), $S_2$ (red dashed curves) and $S_3$ (green dashed-dotted curves). Two different temperatures were used for the calculations in order to take the lowest and the highest measured surface temperature of the dust layers into account, $300 \, \mathrm{K}$ and $450 \, \mathrm{K}$. The straight lines are showing the calculated heat conductivity of the packed-aggregate structures without radiative heat transport. The model predictions, including radiative heat transport, are visualized by the curves. Note that the experimental results can only be explained by the model, if radiative heat transport is taken into account.}
\label{Model_WL}
\end{figure}

The model for the heat conductivity of dust layers composed of individual dust aggregates was tested by a comparison with experimental results obtained by \citet{Krause2011}. In this work, the heat conductivity of dust layers consisting of individual aggregates was investigated experimentally (their samples S1, S2 and S3, see Tab. \ref{Table_Agglomerates}). The dust layers were prepared by sieving the dust through metal filters with different mesh apertures. The mesh aperture determines the maximum size (diameter) of the aggregates. During these experiments, the surface temperature of the dust layer was between $300 \, \mathrm{K}$ and $450 \, \mathrm{K}$. The experimental results are shown in Fig. \ref{Model_WL} (symbols). Additionally, the most important properties of the samples and the used model parameters are summarized in Tab. \ref{Table_Agglomerates}.
\begin{table*}[t!]
\begin{center}
    \footnotesize
    \caption{Properties and model parameters for the dust layers used in the experiments conducted by \citet{Krause2011}. The maximum aggregate size is given by the mesh aperture (MA). Additionally, the volume filling factors of the aggregates $\phi_{Agg}$, the volume filling factors packing structures $\phi_{Struc}$, the corresponding mean free path of the photons inside the voids of the material $\Lambda(R)$ (see Eq. \ref{Theory4}) and the coefficients $\xi(r_0,\phi_{Agg}) \cdot r $ and $\xi(R,\phi_{Struc}) \cdot R $ (see Eq. \ref{Theory1} and Eq. \ref{Theory1a}) are shown.}\vspace{1mm}
    \begin{tabular}{lccccccc}
        \bottomrule
        Experiment & MA $\mathrm{[mm]}$ & $\phi_{Agg}$ & $\phi_{Struc}$ & $\Lambda(R,\phi_{Struc})$  $\mathrm{[m]}$ & $\xi(r_0,\phi_{Agg}) \cdot r $ & $\xi(R,\phi_{Struc}) \cdot R $ & Model \\
        \midrule
        $S_1$ (Plus)   &  $0.50$ & $0.45$ & $0.64$ & $0.8 \, R$ & $0.55$ & $1.50$ & Blue Dotted Curves\\
        $S_2$ (Square) &  $0.25$ & $0.45$ & $0.53$ & $1.2 \, R$ & $0.55$ & $0.84$ & Red Dashed Curves\\
        $S_3$ (Cross)  &  $0.15$ & $0.45$ & $0.35$ & $2.5 \, R$ & $0.55$ & $0.33$ & Green Dashed-Dotted Curves \\
        \bottomrule
    \end{tabular}
     \label{Table_Agglomerates}
     \end{center}
\end{table*}
\par
For the calculation of the heat conductivities of the sieved dust layers, we assume that the typical aggregate size is given by the mesh size. Furthermore, we used a Young's modulus of $E_{Agg} \, = \,  8100 \, \mathrm{Pa}$, which was derived by \citet{Weidling2011} for aggregates with $R \, = \, 0.5 \, \mathrm{mm}$. All aggregates, used in the work by \citet{Weidling2011} and in the work by \citet{Krause2011} were composed of micrometer-sized $\mathrm{SiO_2}$ particles with $r_0 \, = \, 0.75 \, \mathrm{\mu m}$. Thus, we also used this value for the calculations.
\par
Fig. \ref{Model_WL} shows the results of the experiments \citep[symbols;][]{Krause2011} and the model (curves). The calculations were performed for two different temperatures, $300 \, \mathrm{K}$ (lowest surface temperature measured during the experiments) and $450 \, \mathrm{K}$ (highest surface temperature measured during the experiments). For the calculation of the mean free path of the photons (see Eq. \ref{Theory4}), we used the volume filling factor of the packing structure of the aggregates. The variation of the aggregate sizes and the volume filling factors between the three samples leads to different coefficients $\xi(R,\phi) \, R $ (see Eq. \ref{Theory8} and Tab. \ref{Table_Agglomerates}).
\par
Thus, we performed individual calculations for each experiment (see Fig. \ref{Model_WL}). For example, the blue dotted curves are showing the results of the model using the parameters of the $\mathrm{S_1}$ sample. The red dashed and the green dashed-dotted curves correspond to the $\mathrm{S_2}$ and the $\mathrm{S_3}$ sample, respectively. The straight lines are visualizing the heat conductivity of the packed-aggregate structures without radiative transport. Neglecting the influence of the radiation, the heat conductivity decreases with increasing radius of the aggregates. Obviously, the heat conductivity of the solid material (straight lines) can not explain the experimental results.
\par
However, taking the radiative heat transport into account (curves) leads to an increase of the heat conductivity for bigger aggregates. If the radiative heat transport is dominant, the heat conductivity strongly increases with the radius of the aggregates. Hence, a comparison of our calculations with the experimental data shows that the model can reproduce the experimental results within the uncertainty of the experiments (i. e. unknown mean temperature of the dust layers and experimental uncertainties, shown by the error bars.), if radiative heat transport is considered. However, the measured heat conductivities are showing a systematic trend relative to the model. The result of the $\mathrm{S_2}$ experiment fits quite well to the model (red dashed curves), while the result of the $\mathrm{S_1}$ experiment can only be explained with the calculation for $450 \, \mathrm{K}$ (upper blue dotted curve) and the result of the $\mathrm{S_3}$ experiment is only in a good agreement with the model for $300 \, \mathrm{K}$ (green dashed-dotted curves). This can qualitatively be explained by an increasing deviation from a linear temperature gradient from sample S3 through sample S1, due to an increased contribution of radiative heat transport (see also Sect. \ref{Influence on the modeling of cometary activity} and Fig. \ref{CometFig1}). The average temperature throughout the dust-aggregate layer increases from sample S3 through sample S1 even if the bottom and surface temperatures are kept constant, which formally yields the tendency seen in Fig. \ref{Model_WL}.
\par
It is important to note that the model has no free parameter to fit the theory to the experiments. The invented model only depends on the physical properties of the material, e. g. the grain size (pore size), the volume filling factor of the packing structure, the volume filling factor of the individual aggregates and the temperature of the dust layer.
\par
Furthermore, the experimental results can only be explained with the model, if radiative heat transport is considered, which consequently implies that radiation plays an important role for the heat transport in porous materials, even for temperatures $T \, = \, 300 \, \mathrm{K}$. As a result, the heat conductivity of a porous material can not arbitrarily be reduced just by increasing the size of the aggregates or by just increasing the size of the monomeres, because of the influence of the radiative heat transport.

\section{Discussion}\label{Discussion}
In the first part of this section, we show how the simple model (see Sect. \ref{A general model for the heat conductivity of dust layers}) affects the Hertz factor. In the second part, we discuss the influence of the two models on the thermophysical modeling of cometary activity. Finally, in the last part of this section, we demonstrate the limitations of the models.

\subsection{Influence on the Hertz factor}\label{Influence on the Hertz factor}
Here, we use the simple model (see Sect. \ref{A general model for the heat conductivity of dust layers}) to estimate the influence of the temperature of the material and of the radius of the particles on the Hertz factor $H(r,T)$. The Hertz factor,
\begin{equation}
H(r,T) \, = \,\frac{\lambda_{Solid}(r,T)}{\lambda_{Bulk}(T)} \, \mathrm{,}
\label{hertz1}
\end{equation}
describes the reduction of the effective cross section of a porous material, due to the porosity of the material (see Sect. \ref{Heat conductivity}). Using Eq. \ref{Theory3} and \ref{hertz1}, the Hertz factor is given by
\begin{equation}
H(r,T) \, = \, \left[\, \frac{9}{4} \,\frac{1 \, - \, \mu^2}{E(T)}  \, \pi \, \gamma(T) \, r^2 \,\right]^{1/3}  \xi(r,\phi) \, \mathrm{.}
\label{hertz2}
\end{equation}
\par
Eq. \ref{hertz2} shows that the Hertz factor depends on the temperature of the material and on the radius of the particles.
\par
\begin{figure}[t]
\centering
\includegraphics[angle=180,width=0.5\textwidth]{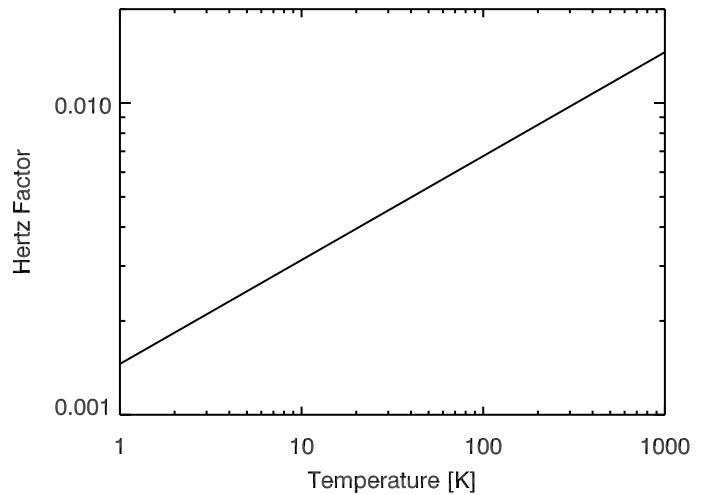}
\caption{Temperature dependence of the Hertz factor (see Eq. \ref{hertz2}) for particles with a radius of $r \, = \, 1 \, \mathrm{\mu m}$ and a volume filling factor of the material of $\phi \, = \, 0.5$.}
\label{Hertz_plot_1}
\end{figure}
\begin{figure}[t]
\centering
\includegraphics[angle=180,width=0.5\textwidth]{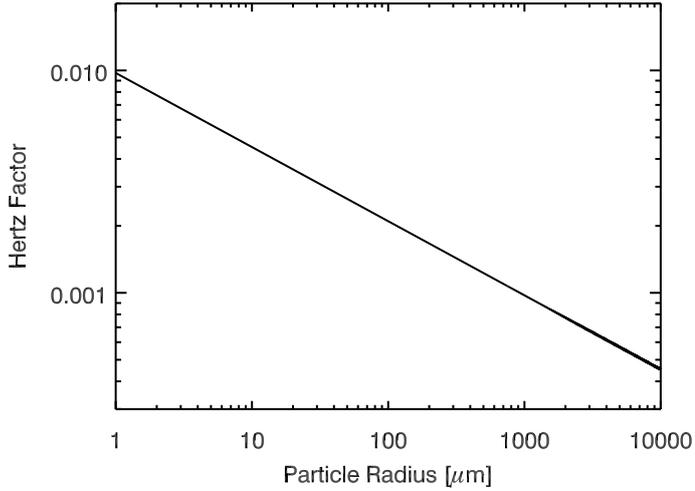}
\caption{Influence of the particle radius on the Hertz factor (see Eq. \ref{hertz2}) for a temperature of $T \, = \, 300 \, \mathrm{K}$ and a volume filling factor of the material of $\phi \, = \, 0.5$.}
\label{Hertz_plot_2}
\end{figure}

Fig. \ref{Hertz_plot_1} shows the temperature dependence of the Hertz factor for particles with a radius of $r \, = \, 1 \, \mathrm{\mu m}$. Additionally, the influence of the grain size on the Hertz factor, for a temperature of $T \, = \, 300 \, \mathrm{K}$, is shown in Fig. \ref{Hertz_plot_2}. In both examples, the volume filling factor of the material is $\phi \, = \, 0.5$.

\subsection{Influence on the modeling of cometary activity}\label{Influence on the modeling of cometary activity}
The absorption of solar radiation by cometary surface layers leads to an increase of the surface temperature. Due to the heat conductivity of the material, heat can be transported into the interior of the cometary nucleus, which can cause sublimation of the icy constituents. While the plane albedo of the surface determines the amount of absorbed energy, the strength of the energy flux through cometary surface layers is given by the heat conductivity of the material and by the vertical temperature gradient inside the material. Modeling the energy distribution in cometary surface layers is a complicated task \citep[see e. g.][]{DavidssonSkorov2002}, which is far beyond the scope of this work. However, we like to discuss the main implications of our model for the heat conductivity of dust layers on the thermophysical modeling of cometary activity.
\begin{figure}[t]
\centering
\includegraphics[angle=0,width=0.5\textwidth]{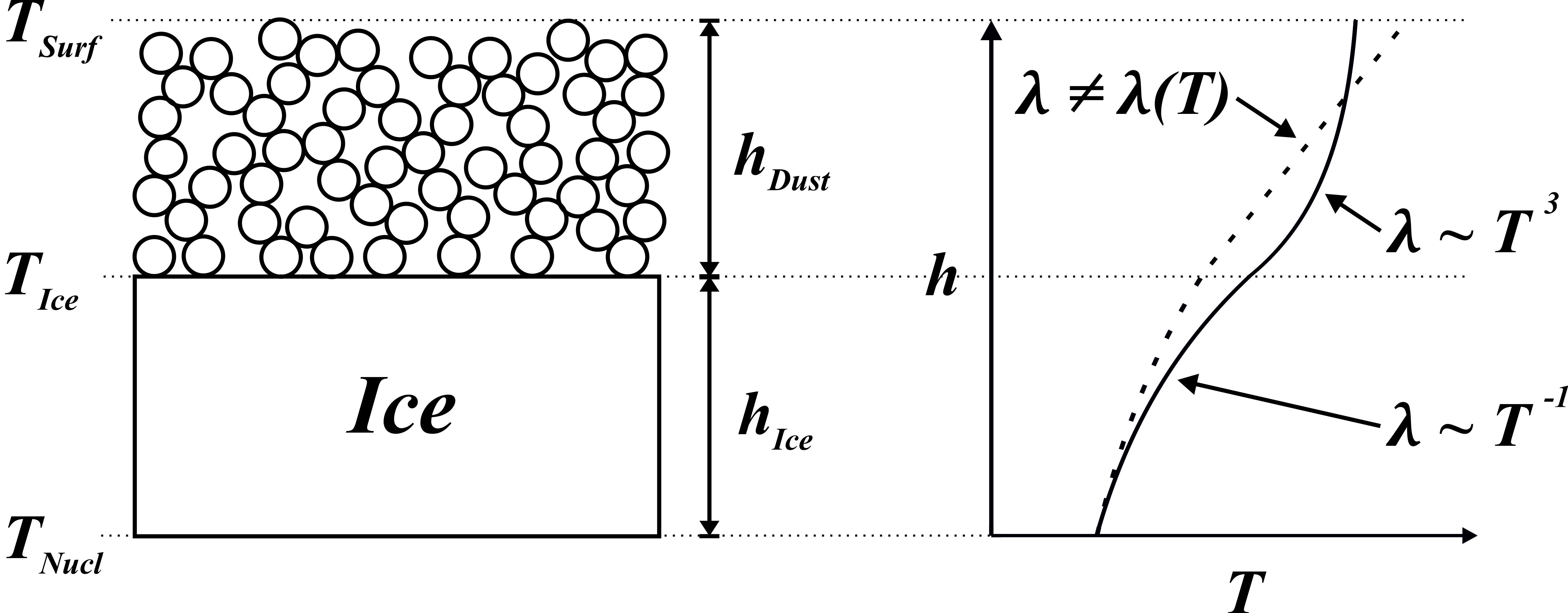}
\caption{Sketch of a cometary surface layer. A solid $\mathrm{H_2O}$ ice layer is covered by a porous dust layer. Additionally, the vertical temperature profiles inside the porous dust layer and inside the solid ice layer are shown, for different heat conductivities, $\lambda \, \neq \, \lambda(T)$ (porous dust layer; see Eq. \ref{Comet2.2}), $\lambda \, \sim \, T^3$ (porous dust layer; see Eq. \ref{Comet4}) and $\lambda \, \sim T^{-1}$ (solid ice layer; see Eq. \ref{Comet6}). In this example, we assume that the heat conduction is dominated by radiation ($\lambda \, \neq \, \lambda(T) \, \ll \, \lambda \, \sim \, T^3$). Furthermore, the temperature of the cometary nucleus $T_{Nucl}$ is fixed, for a better visualization of the influence of temperature dependence of the heat conductivity. Note that the surface temperature of the dust layer $T_{Surf}$ and the temperature of the ice surface $T_{Ice}$ are influenced by the different heat conductivities of the porous dust layer.}
\label{CometFig1}
\end{figure}
\par
In the simplest case, the structure of cometary near-surface regions can be divided into two layers. An ice-dominated layer and an ice-free dust layer, covering the ice (see Fig. \ref{CometFig1}). Then, the surface temperatures of the dust layer and of the ice can be calculated using the energy balance equations at the surface of the dust layer and at the ice-dust interface,
\begin{align}
I \, (1 \, - \, A) \,&=\, \epsilon \, \sigma \, T_{Surf}^4 \, + \, \lambda_{Dust}(r,T,\phi,\Lambda(r)) \, \frac{\partial T(h)}{\partial h}\Bigg|_{Surf}
\label{Comet1}
\end{align}
and
\begin{align}
&\lambda_{Dust}(r,T,\phi,\Lambda(r)) \, \frac{\partial T(h)}{\partial h} \Bigg|_{Ice-Dust}  \,=\, \nonumber\\[2mm]
\lambda_{Ice}(r, & T,\phi,\Lambda(r)) \, \frac{\partial T(h)}{\partial h}\Bigg|_{Ice-Dust}
  + \, Z(T_{Ice}) \, \Delta(T_{Ice}) \, \mathrm{.}
\label{Comet2}
\end{align}
$Z(T_{Ice})$ is the sublimation rate of the ice and $\Delta(T_{Ice})$ is the latent heat of the ice, which determines the cooling of the ice due to the sublimation process. The vertical temperature gradients $\partial T(h) / \partial h$ inside the dust layer and inside the ice can be obtained by solving the heat transfer equation.
\par
Computing the vertical temperature gradient for a constant, i. e. not temperature dependent heat conductivity, $\lambda \neq \lambda(T)$, yields,
\begin{align}
\frac{\partial T(h)}{\partial h}\Bigg|_{\lambda \, \neq \, \lambda(T)} \, = \, \frac{T_{Surf} \, - \, T_{Ice}}{h_{Dust}} \, \mathrm{,}
\label{Comet2.1}
\end{align}
where, $h_{dust}$ is the height of the dust layer. The corresponding vertical temperature profile inside the dust layer is then given by
\begin{align}
T(h) \, \Big|_{\lambda \,\neq \,\lambda(T)}  \, = \, \frac{T_{Surf} \, - \, T_{Ice}}{h_{Dust}} \, h \, + T_{Ice}  \, \mathrm{,}
\label{Comet2.2}
\end{align}
where h=0 is the position of the ice-dust interface. However, the assumption that the heat conductivity does not depend on temperature, is in contradiction with our model for the heat conductivity of dust layers (see Sect. \ref{Theoretical} and Eq. \ref{Theory3}). A temperature dependent heat conductivity leads to a non-linear vertical temperature gradient inside the dust layer. Therewith, the total amount of transported heat through the dust layer is influenced. Thus, a constant heat conductivity should only be used, if the influence of the temperature on the heat conductivity is negligible, e. g. if radiative heat transport is not dominant or if the temperature differences across the dust layer are small (i. e. in our case for $r \, \lesssim \, 15 \, \mathrm{\mu m}$ and $T \, \lesssim \, 300 \, \mathrm{K}$; see Fig. \ref{HeatCondSurfTemp} and \ref{HeatCondPartSize}). For the general calculation of the heat transport through porous dust layers, we propose to use the temperature dependent heat conductivity, introduced in Sect. \ref{Theoretical}.
\par
If the radiative heat transport is dominant, the temperature dependence of the heat conductivity is given by, $\lambda(T) \, \sim \,T^3$ (see Eq. \ref{HeatCond2}). In this case, the solution of the heat transfer equation yields for the vertical temperature gradient inside the dust layer,
\begin{align}
\frac{\partial T(h)}{\partial h}\Bigg|_{\lambda \, \sim \, T^3} \, = \ \ &\frac{1}{4 \, h_{Dust}} \, \left(\, T^4_{Surf} \, - \, T^4_{Ice}\,\right) \nonumber\\[2mm]
&\times\left[\, \frac{h}{h_{Dust}} \, \left( \, T^4_{Surf} \, - \, T^4_{Ice} \,\right) \, +\, T_{Ice}^4 \,\right]^{-3/4}   \, \mathrm{.}
\label{Comet3}
\end{align}
Therewith, we can calculate the vertical temperature profile inside the dust layer,
\begin{align}
T(h)\,\Big|_{\lambda \, = \, \sim \, T^3} \, = \,\left[\, \frac{h}{h_{Dust}} \, \left( \, T^4_{Surf} \, - \, T^4_{Ice} \,\right) \, +\, T_{Ice}^4 \,\right]^{1/4}\, \mathrm{.}
\label{Comet4}
\end{align}
\par
The heat conductivity of solid $\mathrm{H_2O}$ ice also depends on temperature, $\lambda_{Ice}(T) \, = \, g_1 \, T^{-1}$, with $g_1 \, = \, 567 \, \mathrm{W \, m^{-1}}$ \citep{Klinger1980}. Thus, the vertical temperature gradient and the vertical temperature profile inside the ice layer are given by
\begin{align}
\frac{\partial T(h)}{\partial h}\Bigg|_{\lambda \, \sim \, T^{-1}} \, = \, \ &  \ \frac{1}{h_{Ice}}  \ \mathrm{ln}\left(\, \frac{T_{Ice}}{T_{Nucl}}  \,\right)  \nonumber\\[2mm]
& \times \, T_{Nucl} \  \mathrm{exp}\left[\, \frac{h}{h_{Ice}} \  \mathrm{ln}\left(\, \frac{T_{Ice}}{T_{Nucl}}   \,\right) \,\right] \, \mathrm{,}
\label{Comet5}
\end{align}
and
\begin{align}
T(h)\,\Big|_{\lambda \, = \, \sim \, T^{-1}} \, = \, T_{Nucl} \  \mathrm{exp}\left[\, \frac{h}{h_{Ice}} \  \mathrm{ln}\left(\, \frac{T_{Ice}}{T_{Nucl}}  \,\right) \,\right] \, \mathrm{,}
\label{Comet6}
\end{align}
respectively.
\par
The influence of the different heat conductivities on the temperature profiles are shown in Fig. \ref{CometFig1}. In this example, we assumed $\lambda \, \neq \, \lambda(T) \, \ll \, \lambda \, \sim \, T^3$ (i. e. heat conduction is dominated by radiation). Thus, the amount of transported heat through the dust layer is increased from situation I ($\lambda \, \neq \, \lambda(T)$) to situation II ($\lambda \, \sim \, T^3$). This leads to an increase of the ice temperature and to a decrease of surface temperature of the dust layer from situation I to situation II.
\par
$T_{Nucl}$ denotes the temperature inside the nucleus, which is not known in detail. However, the internal temperature of cometary nuclei should be very low. One observational evidence is the presence of $\mathrm{CO}$ around the nucleus of Jupiter-Family Comets, which is interpreted as an indicator for very low temperatures, $T_{Nucl} \, < 50 \, \mathrm{K}$ \citep{Thomas2009}, because the sublimation temperature of CO ice is $T \,\thickapprox\, 26 \, \mathrm{K}$. Note that this interpretation can be misleading, because CO can also be trapped in the form clathrates inside other volatile components.
\par
It was shown that the usage of a temperature dependent heat conductivity influences the vertical temperature gradient inside the dust layer and, therewith, the amount of transported heat through the dust layer. Thus, the usage of a temperature dependent heat conductivity for the calculation of the heat transport inside porous dust layers is crucial.
\par
Furthermore, the heat conductivity of a porous material can not be arbitrarily decreased by increasing the particle size, or by increasing the aggregate size (see Fig. \ref{HeatCondPartSize} and Fig. \ref{Model_WL}). Increasing the particle size, or aggregate size, automatically leads to an enhancement of the heat transport due to radiation. Consequently, the energy transfer on the ice surface can even be enhanced if the particle size is increased.

\subsection{Limitations of the model}\label{Limitations of the model}
An underlying assumption of the model described in Sect. \ref{Theoretical} is that the contacts between the particles or aggregates are formed by van der Waals forces. Thus, our model for the conductive heat transport in porous materials is limited to those regions in which the gravitational compression is negligible with respect to the adhesive force. The hydrostatic force per particle in the simple model is given by
\begin{equation}
F_{Hydr}(r,h)\, = \, g \, \rho \, \pi \, r^2 \, h \,\mathrm{,}
\label{limits1}
\end{equation}
with $g$ and $\rho$ being the local gravitational acceleration and the mass density of the solid material, respectively. This compressive force must not exceed the van der Waals force (see Eq. \ref{Theory2}), if our model shall be applicable.
\par
Similarly, in the complex model, we get for the hydrostatic force per dust aggregate
\begin{equation}
F_{Hydr,Agg}(R,h,\phi_{AL},\phi_{DL})\, = \, g \, \rho \, \pi \, R^2 \, h \, \frac{\phi_{Surf}}{\phi_{Struc}} \, \mathrm{,}
\label{limits2}
\end{equation}
with $\phi_{Surf} \, = \, \rho_{Surf} \, / \, \rho$. Here, $\rho_{Surf}$ is the mass density of the surface material. Following \citet{Weidling2011}, we get for the adhesive force per aggregate-aggregate contact
\begin{equation}
F_{vdW,Agg}(R,T)\, = \, 3 \, \pi \, \gamma_{Agg}(T) \, R \, \mathrm{,}
\label{limits3}
\end{equation}
where $\gamma_{Agg}(T)$ is given by Eq. \ref{Theory5.1} \citep[see also Eq. 5 in][]{Weidling2011}. With suitable material parameters, we get $\gamma_{Agg} \, = \, 2 \times 10^{-5} \, \mathrm{J \, m^{-2}}$.
\par
Using the above equations, we see that for the simple model and suitable $g$ values for comets, $F_{vdW}(r,T)$ always exceeds $F_{Hydr}(r,h)$ so that our simple model (see Sect. \ref{A general model for the heat conductivity of dust layers}) is widely useable. For example, $F_{vdW}(r,T)\, / \, F_{Hydr}(r,h) \, = \, 2 \times 10^2$, for  $g \, = \, 3 \times 10^{-4} \, \mathrm{m \, s^{-2}}$, $r \, = \, 1 \, \mathrm{mm}$, $\rho \, = \, 1000 \, \mathrm{kg \, m^{-3}}$, $h \, = \, 1 \, \mathrm{m}$ and $\gamma \, = \, 0.02 \, \mathrm{J \, m^{-2}}$.
\par
However, for the complex model (see Sect. \ref{Heat conductivity of dust layers composed of individual dust aggregates}), the hydrostatic force exceeds the adhesive force for heights $h\, > \, 0.5 \, \mathrm{m}$ (for $\rho \, = \, 1000 \, \mathrm{kg \, m^{-3}}$, $g \, = \, 3 \times 10^{-4} \, \mathrm{m \, s^{-2}}$, $R \, = \, 1 \, \mathrm{mm}$, $\phi_{Surf} \, = \, 0.2$ and $\phi_{Struc} \, = \, 0.5$). Thus, our expressions for the conductive heat transfer in aggregates of dust aggregates is only valid for shallow depths. However, if the heat transport is dominated by radiative transfer, our model is always applicable.

\section{Conclusion}\label{Conclusion}
In this paper, we investigated the heat transport in porous dust layers in vacuum. Therefore, the heat conductivity of $\mathrm{SiO_2}$ particles was measured experimentally (see Sect. \ref{Experimental technique} - \ref{Results}). Furthermore, we developed a model for the heat conductivity in loose granular media, based on the theory of \citet[][see Sect. \ref{Theoretical}]{ChanTien1973}.
\par
This model was compared with the experimental results. We found that the model predictions are in a good agreement with the experimental results. Note that the model for the heat conductivity of porous dust layers has no free parameter to fit the obtained data. The model for the heat conductivity is capable to distinguish between two different types of dust layers: dust layers composed of single particles (simple model; see Sect. \ref{A general model for the heat conductivity of dust layers}) and dust layers consisting of individual aggregates (complex model; see Sect. \ref{Heat conductivity of dust layers composed of individual dust aggregates}).
\par
The heat conductivity of porous materials in vacuum is determined by two transport mechanisms, the heat conduction through the solid material and the heat transport due to radiation within the pores of the material. The heat transport through the solid material decreases with increasing particle size. The radiative heat transport strongly depends on the temperature of the dust layer and becomes important if the temperature of the dust layer is high (i. e. in our case for $r \, \gtrsim \, 15 \, \mathrm{\mu m}$ and $T \, \gtrsim \, 300 \, \mathrm{K}$; see Fig. \ref{HeatCondSurfTemp} and \ref{HeatCondPartSize}). Additionally, both heat transport mechanisms depend on the packing structure of the material.
\par
In Sect. \ref{Discussion}, we discussed the influence of the model on the Hertz factor. We found that the Hertz factor increases with temperature for fixed particle radius, due to an increase in the van der Waals strength. Furthermore, the Hertz factor decreases with increasing particle radius. Additionally, we discussed the implications of the model for the heat conductivity of porous dust layers on the thermophysical modeling of cometary activity. We have shown that the temperature gradient inside the dust layer and, thus, the transported energy through the dust layer changes if the heat conductivity is temperature dependent (see Sect.  \ref{Influence on the modeling of cometary activity}). Thus, the exact determination of the properties of the dust layer, e. g. the particle size, the temperature and the volume filling factor, is important for the modeling of the energy fluxes inside cometary surface layers.
\par
Finally, we reviewed the limits of the model in Sect. \ref{Limitations of the model}. Both models, the simple model (see Sect. \ref{A general model for the heat conductivity of dust layers}) and the complex model (see Sect. \ref{Heat conductivity of dust layers composed of individual dust aggregates}), are useable, if $F_{vdW}(r,T)$ exceeds $F_{Hydr}(r,h)$. Using this condition, we found that the simple model is widely usable for suitable $g$ values for comets. Furthermore, the complex model is also valid, if the height of the dust layer does not exceed $h \, = \, 0.5 \, \mathrm{m}$.

\subsection*{Acknowledgements}
We thank Steffen Mühle, Kristine Wolling, Maya Krause and Yuri Skorov for the experimental measurements of the heat conductivity and the plane albedo of the dust samples. We also thank Björn Davidsson and an anonymous referee for valuable comments on the manuscript.

\bibliographystyle{model2-names}
\bibliography{bib}

\end{document}